\documentclass[twocolumn]{aa}
\usepackage{graphicx}
\usepackage{txfonts}
\usepackage{mathrsfs} 
\usepackage{hyperref}
\usepackage{xcolor}
\usepackage{natbib}
\usepackage{placeins}
\usepackage[normalem]{ulem}
\usepackage{stfloats}
\usepackage{subcaption}

\makeatletter
\renewcommand*\aa@pageof{, page \thepage{} of \pageref*{LastPage}}
\makeatother

\usepackage[authormarkup=none]{changes} 
\definechangesauthor[name={SF}, color=teal]{sf}

\newcommand{\met}[1]{\text{\tiny#1}}

\begin{document}

\title{Characterizing current structures in 3D hybrid-kinetic simulations of plasma turbulence}
\author{M. Sisti\inst{1,2}\fnmsep\thanks{E-mail: manuela.sisti@univ-amu.fr}\and S. Fadanelli\inst{2}\and S.S. Cerri\inst{3} \and M. Faganello\inst{1}\and F. Califano\inst{2}\and O. Agullo\inst{1}}
\institute{Aix-Marseille University, CNRS, PIIM UMR 7345, Marseille, France, EU\and Dipartimento di Fisica, Universit\`a di Pisa, Italy, EU\and Department of Astrophysical Sciences, Princeton University, 4 Ivy Lane, Princeton, NJ 08544, USA
           }   
\date{Received Month DD, YYYY; accepted Month DD, YYYY}

\abstract
    {In space and astrophysical plasmas turbulence leads to the development of coherent structures characterized by a strong current density and important magnetic shears.} 
    {Using hybrid-kinetic simulations of turbulence (3D with different energy injection scales) we investigate the development of these coherent structures and characterize their shape.} 
    {First, we present different methods to estimate the overall shape of the 3D structure using local measurements, foreseeing application on satellite data. Then we study the local magnetic configuration inside and outside current peak regions, comparing the statistics in the two cases. Last, we compare the statistical properties of the local configuration obtained in simulations with the ones obtained analyzing an MMS dataset having similar plasma parameters.}
    {Thanks to our analysis, 1) we validate the possibility to study the overall shape of 3D structures using local methods, 2) we provide an overview of local magnetic configuration emerging in different turbulent regimes, 3) we show that our 3D-3V simulations can reproduce the structures that emerge in MMS data for the periods studied by \cite{phan_electron_2018, Stawarz_2019}.}
    {}
    

\maketitle

\section{Introduction}

Turbulent, magnetised plasmas permeate a wide range of space and astrophysical environments, and plasma turbulence naturally develops {\it coherent structures} characterized by high current density and strong magnetic shear. These features are indeed present in practically any turbulence simulation employing the most disparate plasma models and regimes \cite[e.g.,][and references therein]{zhdankin_statistical_2013,PassotEPJD2014,BanonNavarroPRL2016,ZhdankinPRL2017,CerriFSPAS2019,ComissoSironiAPJ2019}, as well as routinely observed via {\it in-situ} measurements in space plasmas such as the solar wind and the near-Earth environment \cite[e.g.,][and references therein]{PodestaJGRA2017,GrecoSSRv2018,fadanelli_fourspacecraft_2019,PecoraAPJ2019,GingellJGRA2020,KhabarovaSSRv2021}. The characterization of current structures in turbulent plasmas is of particular interest not only because {\it magnetic reconnection} and/or different {\it dissipation processes}
can occur inside (or close to) these regions, thus enabling energy conversion and plasma heating
\cite[e.g.,][and references therein]{GoslingPhanAPJ2013,TenBargeHowesAPJ2013,OsmanPRL2014,ZhdankinAPJ2014,BanonNavarroPRL2016,GroseljAPJ2017,MatthaeusAPJ2020,AgudeloRuedaJPP2021}, but also because reconnection processes occurring within such structures can in turn feed back onto turbulence itslef by playing a major role in the scale-to-scale energy transfer~\cite[e.g.,][]{carbone_coherent_1990,cerri_reconnection_2017,loureiro_collisionless_2017,franci_magnetic_2017,MalletJPP2017,CamporealePRL2018,DongPRL2018,VechAPJL2018,papini_can_2019}. 

In order to determine the physical behavior of coherent current structures it is of foremost importance to understand how they manifest within the (turbulent) magnetic-field dynamics. This task can be separated into two main inquiries. 1) On the one hand, one must determine the current structure geometry by defining the three characteristic scale lengths of such structures, usually called \textit{thickness} (the smallest), \textit{width} (the medium one) and \textit{length} (the biggest). Such definition of characteristic lengths was employed, for instance, by \cite{zhdankin_statistical_2013} to characterize the current structures emerging in ``reduced-MHD'' simulations of plasma turbulence. While in a simulation we are always in the conditions to precisely define the geometry of a current structure, this is less obvious when it comes to \textit{in-situ} satellite data. A spacecraft can indeed cross a coherent current structure during its path, but it has no information about the overall geometry of it, and even a multi-spacecraft fleet can only measure the spatial variation of physical fields on scales which are in general much smaller than those of any coherent current structure. 2) On the other hand, it is also of key importance to understand how (and if) local magnetic-field configurations within the above mentioned current structures are systematically different from the magnetic configuration that belong to the rest of the (turbulent) environment. While such a characterization can be achieved by a number of procedures, from now on we will focus on the ``Magnetic Configuration Analysis'' (MCA) method proposed by \citet{fadanelli_fourspacecraft_2019}. The MCA method consists of a modification of existing techniques that have been previously employed to investigate the local configuration of magnetic field (namely, the Magnetic Directional Derivative by \citet{shi_dimensional_2005} and the Magnetic Rotational Analysis by \cite{shen_new_2007}; see \citet{shi_dimensionality_2019} for a review of these different techniques). Contrary to measures of current structure geometry, the analysis of magnetic configurations can be performed on data from plasma simulations as well as on data coming from multi-spacecraft missions. For instance, in \citet{fadanelli_fourspacecraft_2019} the authors apply the MCA technique on long intervals of data collected by the Magnetospheric Multiscale (MMS) mission (see \cite{burch_magnetospheric_2016}) determining that it is possible to obtain statistics of local configurations developing in the outer magnetosphere, in the magnetosheath, and in the near-Earth solar wind.

In the first part of this work, we develop and describe three methods by which it is possible to estimate some aspects of a current structure geometry starting from local measurements. We test these methods on the structures emerging in 3D-3V Hybrid-Vlasov-Maxwell (HVM) simulations of plasma turbulence (three spatial dimensions plus three-dimensional velocity space, with kinetic ions and fluid electrons). By comparing the results obtained from these three local methods with those resulting from a non-local approach, we show that it is indeed possible to estimate the overall shape of a current structure by employing using only local measurements. 

The second part of this work investigates the physical characteristics of current peak regions forming in three different three-dimensional Hybrid-Vlasov (3D-3V HVM) simulations. 
In particular, we analyze the different features of magnetic configurations in the three different simulations and we make a comparison between them and those obtained analyzing the MMS observational data of plasma turbulence measured on December 9th 2016 \cite{phan_electron_2018,Stawarz_2019}.  
Indeed, one simulation setup that we include in this work is a three-dimensional equivalent of the 2D setup employed by \cite{Califano_2020}, which proved capable of qualitatively reproducing the turbulent and reconnection regime observed during that period.

This paper is organized as follows.   
In Sections \ref{sec:simulation} and \ref{sec:spectral_analysis} we present the main features of the HVM simulations of plasma turbulence that are employed here, including an overview of the turbulent spectra that we obtain in the different cases. 
In Section \ref{sec:definitions} we give a precise definition of ``current structure'' in our simulations and define two non-local/overall shape factors called planarity $\mathcal{P}$ and elongation $\mathcal{E}$ to better highlight the 3D-shape of a current structure. In the same Section we clarify what we mean by ``magnetic configuration'' and how MCA defines the two shape factors planarity $\mathscr{P}$ and elongation $\mathscr{E}$ to characterize any such configuration. Then, in Section \ref{sec:shape-current-structures} we present three different methods by which it is possible to convert purely local measures into an estimate of $\mathcal{E}$ and $\mathcal{P}$ of current structures, and show their effectiveness. In Section \ref{sec:results} we investigate the physical features of the current regions forming in the 3D simulations we make use here and perform, using our simulations, the analysis proposed by \cite{fadanelli_fourspacecraft_2019}. By performing  this analysis we can show that magnetic configurations inside the coherent current regions behave differently with respect to those in the rest of plasma. Moreover, we show how the statistical distributions of these quantities obtained by numerical simulations well reproduce the ones obtained by analyzing MMS data of December 9th 2016 \cite{phan_electron_2018,Stawarz_2019}. Finally, we summarize the results obtained and their importance in Section \ref{sec:conclusions}.  


\section{Simulations}\label{sec:simulation}
In this paper to investigate the emergence of (coherent) current structures and characterize their nature, we make use of kinetic numerical simulations of plasma turbulence performed with HVM model with kinetic ions and fluid neutralizing electrons with mass \cite{valentini_hybrid-vlasov_2007}.
The corresponding set of equations is normalized using the ion mass $m_i$, the ion cyclotron frequency $\Omega_{ci}$, the Alfv\'en velocity $v_{_A}$ and the ion skin depth $d_i = v_{_A}\Omega_{ci}^{-1}$. As a result, the dimensionless electron skin depth is given by the electron-to-ion mass ratio, $d_{\rm e} = \sqrt{m_{\rm e}/m_{\rm i}}$.

The ion distribution function $f_i=f_i({\bf x},{\bf v},t)$ evolves following the Vlasov equation that in dimensionless units reads
\begin{equation}{\label{eq:vlasov}}
\frac{\partial f_i}{\partial t}+{\bf v}\cdot\nabla f_i + ({\bf E}+ {\bf v}\times{\bf B})\cdot\frac{\partial f_i}{\partial {\bf v}}=0\,
\end{equation}
giving the number density $n$ and the ion fluid velocity ${\bf u}_{\rm i}$ as moments of $f_i$. Then, the electron fluid response is given by a generalized Ohm's law for the electric field $\mathbf{E}$ including the Hall, diamagnetic and electron-inertia effects:
\begin{equation}{\label{eq:ohm}}
\left(1-d_e^2\nabla^2\right){\bf E}\,
=\,
-{\bf u}_{\rm i}\times{\bf B}\,
+\frac{{\bf J}\times{\bf B}}{n}\,
-\frac{\nabla P_{\rm e}}{n}\,
+\frac{d_{\rm e}^2}{n}\nabla\cdot\left( 
{\bf u}_{\rm i}{\bf J}+{\bf J}{\bf u}_{\rm i}-\frac{{\bf JJ}}{n}\right)\,
\end{equation}
In the electron inertia term we approximate $1/n = 1$, in normalized units, for the sake of computational simplicity. 
Furthermore, ${\bf J}=\nabla \times {\bf B}$ is the current density (neglecting the displacement current in the low-frequency regime). 
In equation \eqref{eq:ohm} we assume an isothermal equation of state for the electron pressure, $P_e = nT_{0e}$, with a given initial electron-to-ion temperature ratio $T_{0e}/T_{0i}=1$. The initial ion distribution function is given by a Maxwellian distribution with corresponding uniform temperature. 
Finally, the evolution of the magnetic field $\mathbf{B}$ is given by the Faraday equation:
\begin{equation}{\label{eq:faraday}}
\frac{\partial {\bf B}}{\partial t}=-\nabla\times{\bf E}\,.
\end{equation}

We take an initial magnetic field given by a uniform background field, $B_0{\bf e_z}$ where $B_0 = 1$, with a superimposed small-amplitude 3D perturbation, $\delta {\bf B} = \delta B_x{ \bf e_x} + \delta B_y {\bf e_y} + \delta B_z {\bf e_z}$, computed as the curl of the vector potential, $\delta {\bf B} = \nabla \times \delta {\bf A}$. In particular, $\delta {\bf A}$ is given by a sum of sinusoidal modes with random phase in a limited wavevector interval corresponding to the largest wavelengths admitted by the numerical box. In Table \ref{tab:table1} we list the simulation parameters of four different simulations as follows: the box spatial size, the corresponding number of grid points, the wave vector range of the initial perturbation, the root mean square amplitude of each component of the perturbed magnetic field and the ratio between the root mean square of the magnetic perturbation and the equilibrium field. We sample for all simulations the velocity space using $51^3$ uniformly distributed grid point spanning $[-5v_{th,i}, 5v_{th,i}]$ in each direction, where $v_{th,i} = \sqrt{\beta_{i}/2\,}\,v_A$ is the initial ion thermal velocity and $\beta_{i}=1$. We set a reduced mass ratio $m_i/m_e = 100$.

In the following we will refer to the 3D-3V simulations (the first three listed in Table \ref{tab:table1}) using the following names: ``SIM A", ``SIM B-wf" (weak forcing scenario) and ``SIM B-sf" (strong forcing scenario). We also make use as a reference for comparison of a 2D-3V hybrid Vlasov-Maxwell simulation, namely ``SIM 2D'' (last one in Table \ref{tab:table1}).  
\begin{table*}
\caption{\label{tab:table1} Simulation parameters}
\centering
\begin{tabular}{ccccc}
& \textbf{SIM A} &\textbf{SIM B-wf} & \textbf{SIM B-sf} & \textbf{SIM 2D} \\
      \hline
     Cubic box & $(10*10*10)2\pi d_i$ & $(3*3*3)2\pi d_i$ & $(3*3*3)2\pi d_i$ & $(50*50*1)2\pi d_i$ \\
     Grid points & $352*352*198$ & $256*256*256$ & $256*256*256$  & $3072*3072*1$ \\
     $k_\perp d_i$ range & [$0.1$, $17.6$] & [$0.33$, $42.7$] & [$0.33$, $42.7$]  & [$0.02$, $30.7$] \\
     $k_z d_i$ range & [$0.1$, $9.9$] & [$0.33$, $42.7$] & [$0.33$, $42.7$]  & $\{0\}$ \\
     Energy injection scale & $ 0.1 \leq kd_i < 0.33 $ & $ 0.33 \leq kd_i < 1 $ & $ 0.33 \leq kd_i < 1 $ & $ 0.02 \leq kd_i < 0.12 $\\
     $\big(\delta B^{\rm(rms)}/B_0\big)_{t=0}$ & 0.28 & 0.20  & 0.50 & 0.28 \\
     $\big(\delta B_{z}^{\rm(rms)}/\delta B_{\perp}^{\rm(rms)}\big)_{t=0}$ & 0.7 & 0.75 & 0.75 & 0.95 \\
\end{tabular}
\end{table*}
The reasons behind the choice of simulation parameters are discussed in the following section, providing also a brief overview of turbulence evolution and properties.


\section{Context: turbulent evolution and spectra}
\label{sec:spectral_analysis}

The simulations considered in this work are meant to reproduce the behavior of a $\beta\approx1$ turbulent plasma, typical of the near-Earth's environment and of the solar wind at about $1$ AU (see e.g., \cite{ChenJPP2016,SahraouiRMPP2020} and references therein). 

As summarized in Table~\ref{tab:table1} each simulation initially injects fluctuations' energy in a different wavenumber range, with different root-mean-square (rms) values of the initial magnetic fluctuations exploring a different amount of sub-ion-gyroradius scales. 
In particular, SIM B-wf and SIM B-sf inject energy only slightly above the ion characteristic scales, and are able to excite a wide range of kinetic scales before reaching their dissipation scale (i.e., resolving about a decade of clean sub-ion range turbulence before entering the dissipation-dominated scales). 
In fact, as done in \cite{Califano_2020} for the 2D-3V case, the aim of both SIM B-wf and SIM-sf is to mimic the conditions possibly encountered in the Earth's magnetosheath, past the bow shock, where the occurrence of electron-only reconnection has been observed \cite{phan_electron_2018,Stawarz_2019}.
For completeness, we include a simulation (SIM A) where energy is injected at larger scales including part of the final MHD turbulent range and a simplified two-dimensional one (SIM 2D) covering an even larger range.

We recognize three phases in the temporal evolution of all simulated systems: (1) an initial phase, (2) a transition phase and (3) a fully developed turbulence. At the beginning (1) the energy injected at large scales starts to cascade towards smaller and smaller scales. In the transition phase (2) coherent structures, i.e. regions where the current density peaks (it usually happens at about one eddy-turnover time) start to form. At the end, a fully developed turbulent state (3) is reached, with a complex nonlinear dynamics with coherent structures continuously forming, merging and/or being destroyed. We will indicate the times at which the 3D simulations reach saturation (maximum of the root mean square of the current density) as $t_{sat_A}$ (SIM A), $t_{sat_{B-wf}}$ (SIM B-wf) and $t_{sat_{B-sf}}$ (SIM B-sf); at these times turbulence is fully-developed (roughly speaking saturation times correspond to about 3 eddy-turnover times). From now on, unless explicitly stated otherwise, all the analysis will be carried out at saturation times.

\begin{figure*}
    \centering
    \includegraphics[width=\textwidth]{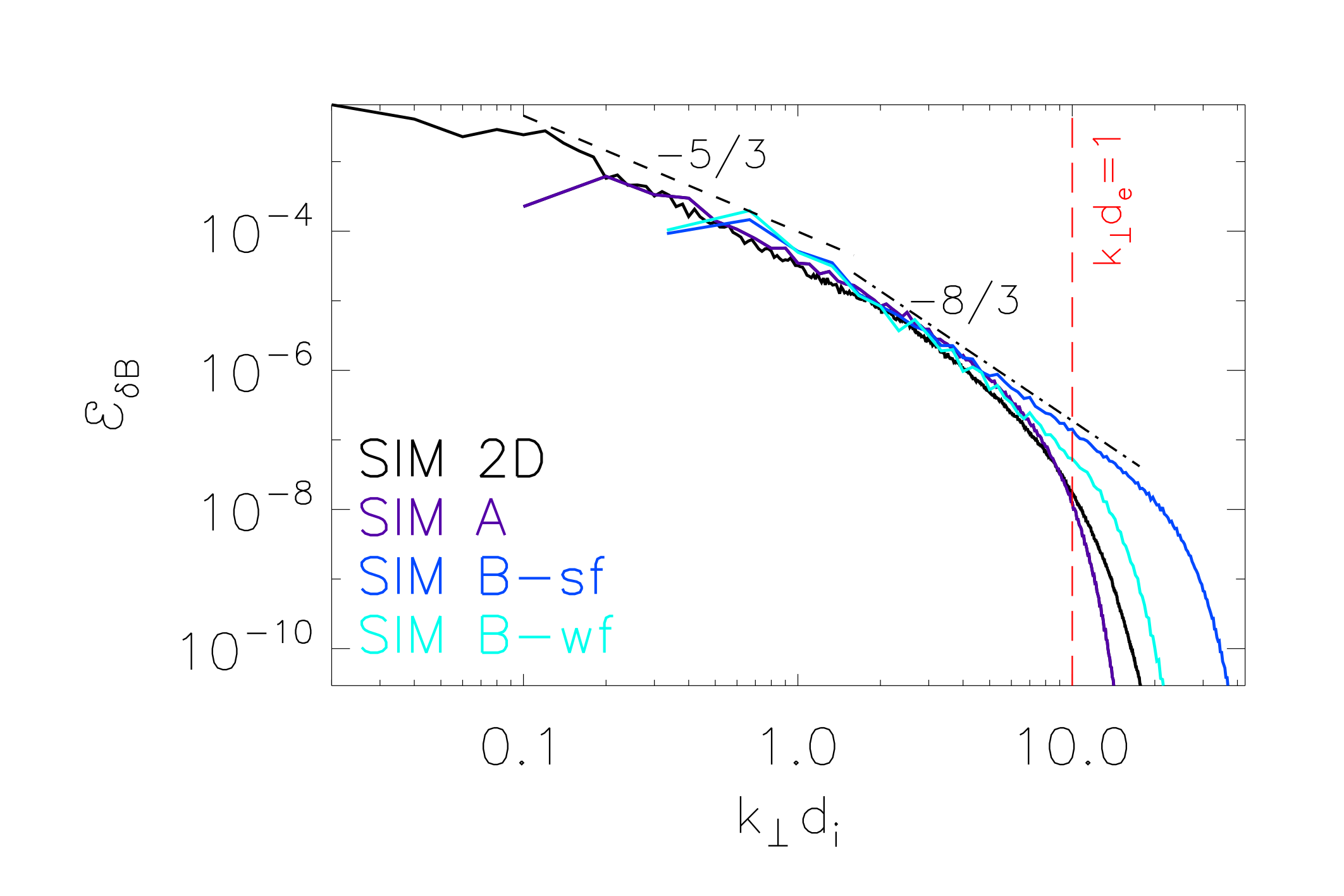}
    \caption{For our four simulations one-dimensional $k_{\parallel}$-averaged magnetic energy spectra versus $k_{\perp}$, normalized so that they overlap at $k_\perp d_i\approx2$.}
    \label{fig:spectra}
\end{figure*}
In Figure \ref{fig:spectra} we show the reduced one-dimensional, $k_{\parallel}$-averaged, magnetic energy spectra versus $k_{\perp}$ for the simulations listed in Table~\ref{tab:table1}. The spectra, for visual purposes, have been normalized so that they overlap at $k_\perp d_i\approx2$. The power laws  $k_\perp^{-5/3}$ and $k_\perp^{-8/3}$ are displayed as references for the MHD and kinetic range, respectively. At ``fluid'' perpendicular scales, in the range $0.1 \le k_{\perp} d_i \le 2$, a power law close to (but slightly steeper than) $-5/3$ is visible in both SIM A and SIM 2D. The spectral slope evaluated at $k_\perp d_i\lesssim1$ for these two simulations (excluding the first modes, which are affected by the initial condition), is indeed in the range $-1.9\lesssim\alpha\lesssim-1.8$ (the actual value slightly depends on the exact wavenumber range on which we perform the fit). On the other hand, SIM B-wf and SIM B-sf do not include enough MHD scales to draw any conclusion regarding the emerging spectral slope for $k_\perp d_i<2$. While a difference between the observed slope in SIM 2D and the expected $-5/3$ power law may be related to the reduced dimension that prevents to properly establish critical balance \cite{SchekochihinAPJS2009}, this discrepancy is also observed in the spectrum emerging from the 3D case (SIM A, whose magnetic-field spectrum seems to agree very well with the corresponding spectrum of SIM 2D). As noticed also in previous 3D-3V simulations \cite{CerriFSPAS2019}, whether this feature in SIM A is due to the limited extent of the MHD range or not is unclear, and will have to await further investigation by means of even larger 3D simulations. 
At ``kinetic'' scales in the range $k_\perp d_i > 2$, a power law fairly consistent with $-8/3$ clearly emerges only in SIM B-sf (a hint of such power law is visible also in SIM A, but the limited extent of the sub-ion range does not allow it to develop over a broad range of scales before the numerical dissipation takes over). Such a slope would be consistent with an ``intermittency corrected'' kinetic-Alfv\'en-wave cascade occurring at sub-ion scales \cite{boldyrev_spectrum_2012}, and previously reported via 3D-3V hybrid-Vlasov simulations \cite{cerri_kinetic_2017,cerri_dual_2018}. The other two simulations, SIM 2D and SIM B-wf, exhibit a power law steeper than $-8/3$. A magnetic-field spectrum close to $k_\perp^{-3}$ in the kinetic range was already observed in previous 2D hybrid-kinetic simulations (see e.g., Ref. \cite{cerri_plasma_2017} and references therein),  sometimes attributed to the effect of a reconnection-mediated two-dimensional cascade at sub-ion scales \cite{cerri_reconnection_2017,franci_magnetic_2017} (which, however, may continue to hold also in three spatial dimensions \cite{loureiro_collisionless_2017, MalletJPP2017}, although a definitive evidence of such regime has been elusive in 3D kinetic simulations performed thus far \cite{CerriFSPAS2019}; see also \cite{AgudeloRuedaJPP2021} for a more recent attempt). For what concerns SIM B-wf, on the other hand, a steeper spectrum may be attributed to a weaker cascade associated to the lower amplitude (with respect to SIM B-sf) of the injected fluctuations, rather than to dissipation effects associated to ion-heating and/or fluctuations' damping mechanisms in the $\beta\approx1$ regime considered here (see e.g., Refs. \cite{ToldPRL2015,SulemAPJ2016,ArzamasskiyAPJ2019} and references therein). In fact, in SIM B-wf there is no evidence of a clear cascade developing along the direction parallel to ${\bf B}_0$ (not shown here). However, a detailed analysis of fluctuations' spectral properties and associated ion-heating mechanisms (and how they change between different simulations) is beyond the scope of this work and will be reported elsewhere. 

\begin{figure*}
    \centering
    \includegraphics[width=0.8\textwidth]{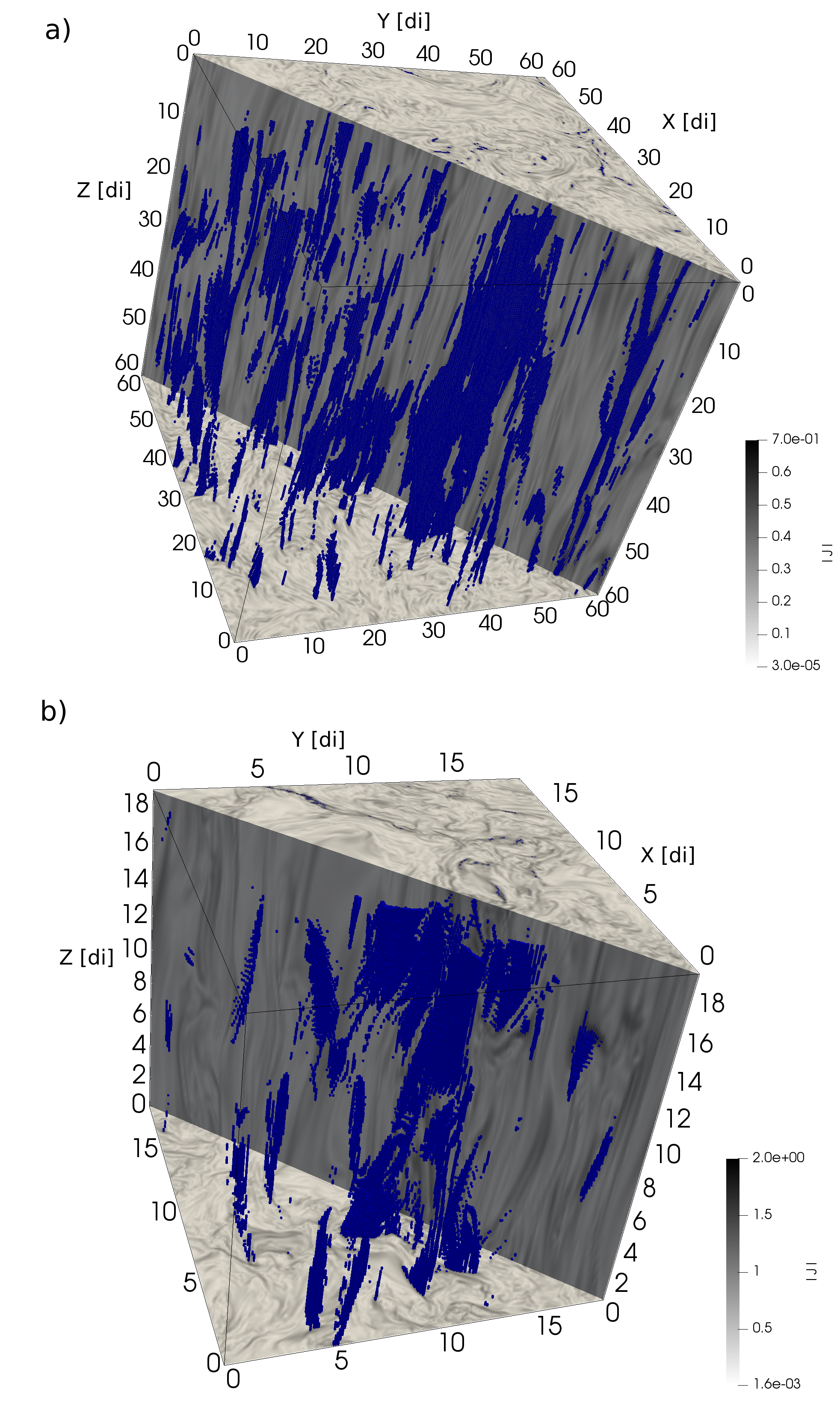}
    \caption{3D rendering of regions with $J\ge J_{th}$ in blue for a) SIM A, b) SIM B-st. In grey, shadow colour plots of current density.}
    \label{fig:3D:view}
\end{figure*}

In Figure \ref{fig:3D:view} we show for a) SIM A, b) SIM B-st, the 3D rendering of the current density magnitude in shaded isocontours (we remind the reader that SIM B-st employs a simulation box size that, compared to the domain of SIM A, is about one third in each direction). 
In blue the regions of grid points which overcome a specific threshold, defined in the next Section, on the current density $J_{th}$.  

\section{Current structures and magnetic configuration analysis}

\subsection{Definitions}\label{sec:definitions}
In order to analyze the physics of current structures, we first  define the procedure aimed at identifying a current structure as well as the parameters to be adopted in order to characterize their shape. Alongside with such definitions, we will also discuss the concept of ``magnetic configuration'' and the meaning of its shape, a concept which will be thoroughly applied in the following sections.

Similarly to what is done in \cite{uritsky_2010,zhdankin_statistical_2013}, we adopt the formal definition of ``current structure'' as any connected region where the current-density magnitude, $J\equiv|{\bf J}|$, is above a threshold value for all internal points, and the boundaries do not touch any other such zone. In particular, we stress that the specific definition of such threshold may be somewhat arbitrary. In this work, consistently with \cite{uritsky_2010} and \cite{zhdankin_statistical_2013}, we choose a current threshold defined by $J_{th}=\sqrt{\langle J^2 \rangle+3\sigma(J^2)}$, where $\sigma(J^2)=\sqrt{ \langle J^4 \rangle -( \langle J^2 \rangle )^2}$ and $\langle ... \rangle$ denotes spatial average over the whole simulation box. We want to remark that, by this procedure, any isolated point with above-threshold current density is not recognized as a current structure by its own, but it can be identified as part of some other current structure if the two are within a minimum distance of one grid-point from each other. For shorthand, we call ``center'' of a structure the point within the structure domain for which $J$ is maximum.

To characterize the spatial extension (or ``geometry'') of a current structure,  we introduce three characteristic lengths $\ell_{max}$, $\ell_{med}$ and $\ell_{min}$. In particular, the ``length'' $\ell_{max}$ indicates the longest among the structure's dimensions (this goes to infinity in 2D systems), the ``thickness'' $\ell_{min}$ is the smallest dimension, and we call ``width'' $\ell_{med}$ the intermediate characteristic length. 

To determine these quantities, we first compute the eigenvectors of the Hessian matrix of the current density at the center of each structure. Then, we consider the plane perpendicular to the direction of least variation (which is the one associated to $\ell_{max}$). 
In this particular plane we have a direction of strongest variation for the current density (associated to $\ell_{min}$) and a direction of weakest variation (associated to $\ell_{med}$). We define the thickness $\ell_{min}$ as the maximum distance between two points with $J \geq J_{th}$ along the direction of strongest variation. On the contrary the width $\ell_{med}$ is the maximum distance between two points with $J \geq J_{th}$ on the plane. Finally, the length $\ell_{max}$ is defined as the maximal distance between two points belonging to the same structure without restricting to special planes. We note here that the definition of the thickness $\ell_{min}$ is different from that of \cite{zhdankin_statistical_2013} where the use of the full width at half maximum of the interpolated profile of $J$ is preferred. Here we decide to define the thickness using the maximum distance between two points with $J\ge J_{th}$ along the direction of strongest variation, in order to remain coherent with the definition of width and length. In the 2D case we have only thickness and width, computed as above. In Appendix \ref{appendix:alternative_dimensions} we clarify what is the impact of slightly different definitions of $\ell_{min}$ on our analysis. 

\begin{figure*}
    \centering
    \includegraphics[width=0.5\textwidth]{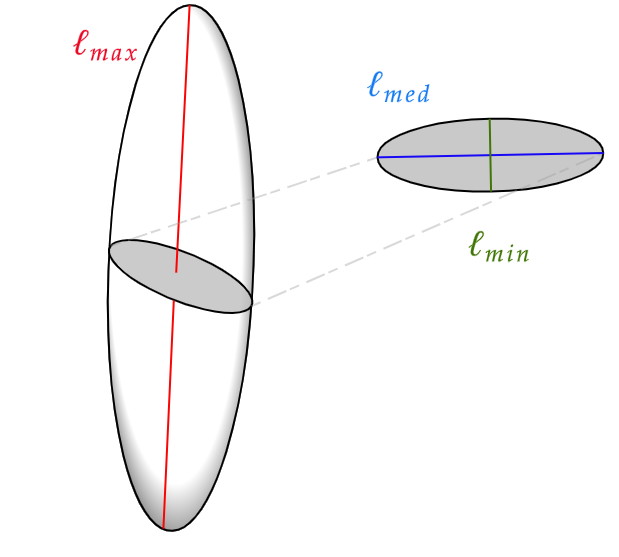}
    \caption{schematic representation of a current structure with its characteristics lengths: $\ell_{min}$, $\ell_{med}$ and $\ell_{max}$.}
    \label{figure:cartoon:structure}
\end{figure*}

Once that the ``geometry'' has been clarified, it is immediate to define also the ``shape'' of current structures. In a current structure, we call ``shape'' the quantity determined by two ``structure shape factors'' defined as:
    \begin{equation}\label{eq:el-str}
    \text{elongation (current str.)} \qquad 
    \mathcal{E} = 1- \ell_{med} / \ell_{max}
     \end{equation}
    \begin{equation}\label{eq:pl-str}
    \text{planarity (current str.)} \qquad 
    \mathcal{P} = 1- \ell_{min} / \ell_{med}
    \end{equation}
These two parameters measure the tendency of the current structure to squeeze toward some elongated form ($\mathcal{E}\rightarrow 1$) or to flatten ($\mathcal{P}\rightarrow 1$). In this way, we can represent each structure's shape in the plane $\mathcal{E}-\mathcal{P}$, and eventually classify shapes into categories. In particular, we adopt the following nomenclature for certain characteristic shapes: (i) ``pseudo-spheres'' (low $\mathcal{E}$, low $\mathcal{P}$), (ii) ``cigars'' (high $\mathcal{E}$, low $\mathcal{P}$), (iii) ``pancakes'' (low $\mathcal{E}$, high $\mathcal{P}$), (iv) ``knife blades'' (high $\mathcal{E}$, high $\mathcal{P}$).
In Figure \ref{figure:cartoon:structure} we show a schematic representation of a current structure with its characteristics lengths: $\ell_{min}$, $\ell_{med}$ and $\ell_{max}$.
We note that the three quantities of thickness, width and length, here presented, are enough to define the shape of compact structures in 3D. In case the holes would appear in current structures, new parameters should be added as a filling fraction or  the hole dimensions. Such upgrade will be investigated in  future work.

By ``magnetic configuration'' we mean the {\it local} characterization of the ${\bf B}$ field that can be inferred at any point in our system, even far from a current structure or its peak, by inspection of the (symmetric) tensor ${\bf N} \equiv [\nabla{\bf B}]\cdot[\nabla{\bf B}]^T/B^2$~\cite{fadanelli_fourspacecraft_2019}. 
Because of its symmetric nature, ${\bf N}$ can be defined in terms of its orthogonal eigenvectors and corresponding eigenvalues: we denote these eigenvalues by $\sigma_{max}$, $\sigma_{med}$, $\sigma_{min}$.
The quantities $1/\sqrt{\sigma_{min}}$, $1/\sqrt{\sigma_{med}}$, $1/\sqrt{\sigma_{max}}$ can be understood as proportional to the three characteristic variation lengths. Here and elsewhere, when considering the eigenvalues, it is important to stress that they are only proportional to the characteristic variation lengths, not directly comparable, due to the presence of normalization factors. What is really significant is the ratio between eigenvalues, as we will see in the next. Indeed, taking the ratio between eigenvalues the normalization factors cancel each other out.

As with current structures, a ``shape'' can be defined for magnetic configurations. The idea is to deal with $1/\sqrt{\sigma_{min}}$, $1/\sqrt{\sigma_{med}}$ and $1/\sqrt{\sigma_{max}}$ in the exact same way we did with characteristic lengths, and define two ``configuration shape factors'' as follows: 
    \begin{equation}\label{eq:el-con}
    \text{elongation (magnetic conf.)} \qquad 
    \mathscr{E} = 1-\sqrt{ \sigma_{min} / \sigma_{med} }
    \end{equation}
    \begin{equation}\label{eq:pl-con}
    \text{planarity (magnetic conf.)} \qquad 
    \mathscr{P} = 1-\sqrt{ \sigma_{med} / \sigma_{max} }
    \end{equation}
We note here that the ``configuration shape factors'' are local quantities, which can be computed at any spatial location, including points owning to a current structure. On the contrary the ``structure shape factors'' provide an overall/non-local picture of a current structure. As $\mathcal{E}$ and $\mathcal{P}$ do for current structures, $\mathscr{E}$ and $\mathscr{P}$ do measure the tendency of magnetic configurations toward elongation (if $\mathscr{E} \to 1$) or squashing into sheets (if $\mathscr{P} \to 1$). Similarly, the position of a configuration in the $\mathscr{E}$-$\mathscr{P}$ plane allows to recognize the local shape of $\bf B$ in one of the following four categories: (i) ``pseudo-sphere'', (ii) ``cigar'', (iii) ``pancake'' or (iv) ``knife-blade''. 

We remind the reader that the work by \cite{fadanelli_fourspacecraft_2019} is focused only on satellite data and thus on local measurements, providing point by point value for $\mathscr{E}$ and $\mathscr{P}$ along spacecrafts' trajectories. On the contrary here, taking advantage of the 3D data of a simulation, we can also give an overall/non-local picture of current structures (via $\mathcal{E}$ and $\mathcal{P}$). 


\subsection{The shape of current structures}\label{sec:shape-current-structures} 

The first goal of this Paper is to determine whether it is possible to estimate the ``non-local'' overall shape of a current structure by {\it local measurements}, i.e. if there is some procedure to convert purely local measurements of certain quantities into a reliable estimate of the ``actual'' elongation and planarity of a given structure. We point out here that we consider the estimate of $\mathcal{E}$ and $\mathcal{P}$ as given by our ``non-local'' overall method in equations \eqref{eq:el-str} and \eqref{eq:pl-str} to be used as ``reference values'' for the estimates obtained via the following {\it local} methods:
\begin{itemize}
    \item {\bf ``HJ'' method.} This approach is based on the Hessian matrix of $J$ calculated at the center of each current structure. If we suppose that ratios of the eigenvalues of this matrix, $h_{max}$, $h_{med}$ and $h_{min}$, can reproduce the ratios between $\ell_{max}$, $\ell_{med}$ and $\ell_{min}$, then the shape factors of the structure can be estimated as: 
        \begin{equation}\label{eq:sha-HJ}
        \begin{aligned}
        \mathcal{E}^{\met{HJ}} &= \left[  1-\sqrt{ h_{min} / h_{med} } \right]_{cen} \\
        \mathcal{P}^{\met{HJ}} &= \left[  1-\sqrt{ h_{med} / h_{max} } \right]_{cen}
        \end{aligned}
        \end{equation}
    where the subscript ``cen'' indicates that the values are calculated at the structure's center. This method, we note, closely resembles the one adopted by \cite{servidio_magnetic_2009}.  
    \item {\bf ``NB'' method.} This approach aims at inferring the ``structure shape factors'', $\mathcal{E}$ and $\mathcal{P}$, from the ``magnetic-configuration shape factors'', $\mathscr{E}$ and $\mathscr{P}$, evaluated through the matrix ${\bf N}$ at the center of a given current structure: 
        \begin{equation}\label{eq:sha-NB}
        \begin{aligned}
        \mathcal{E}^{\met{NB}} &= \left[  1-\sqrt{ \sigma_{min} / \sigma_{med} } \right]_{cen} = \mathscr{E}_{cen} \\
        \mathcal{P}^{\met{NB}} &= \left[  1-\sqrt{ \sigma_{med} / \sigma_{max} } \right]_{cen} = \mathscr{P}_{cen}
        \end{aligned}
        \end{equation}
    where the subscript ``cen'' indicates that the values are calculated at the structure's center only, and $\sigma_{max}$, $\sigma_{med}$, $\sigma_{min}$ are eigenvalues of the tensor $\bf N$ which evaluates the magnetic configuration. it is worth noticing that, while the $\bf N$ tensor is well defined everywhere, for the sake of comparison with the other methods we consider here only the value at the center assuming it as representative for the whole structure. However, to easily make a comparison with satellites data and/or to increase the dataset analyzed, one can  apply this method even to points different from the central one. Since magnetic field and current density are strongly related (in our model, the current density is the curl of the magnetic field), we also expect $\mathcal{E}^{\met{NB}}$ and $\mathcal{P}^{\met{NB}}$ to resemble $\mathcal{E}^{\met{HJ}}$ and $\mathcal{P}^{\met{HJ}}$. 
    \item {\bf ``AV'' method.} This approach considers an average of the magnetic configurations occurring inside each structures. That is, shape parameters are estimated as: 
        \begin{equation}\label{eq:sha-AV}
        \begin{aligned}
        \mathcal{E}^{\met{AV}} &= \left\langle  1-\sqrt{ \sigma_{min} / \sigma_{med} } \, \right\rangle_{str} = \left\langle  \mathscr{E} \, \right\rangle_{str} \\
        \mathcal{P}^{\met{AV}} &= \left\langle   1-\sqrt{ \sigma_{med} / \sigma_{max} } \, \right\rangle_{str} = \left\langle  \mathscr{P} \, \right\rangle_{str}
        \end{aligned}
        \end{equation}    
    where $\langle ... \rangle_{str}$ is the average over all points belonging to a single current structure.  
    This procedure, we note, resembles the NB method but can be carried on without knowing where the center of a current structure is located.  
\end{itemize}


\subsection{Method comparison}\label{subsec:method_comparison}

Our goal here is to verify the accuracy of the different methods HJ, NB and AV in estimating $\mathcal{E}$ and $\mathcal{P}$, that are obtained looking at the overall/non-local shape of the current structure and are thus considered our reference values.  To establish how accurate the different methods detailed in Section \ref{sec:shape-current-structures} are, in this Section we apply them to SIM A, which employs the largest physical domain (and thus is the simulation developing the largest number of current structures to test). To this end, we compute elongation and planarity of every current structure, then compare them with their estimates obtained by the HJ method, NB method and AV method.

\begin{figure*}
    \centering
    \includegraphics[width=0.50\textwidth]{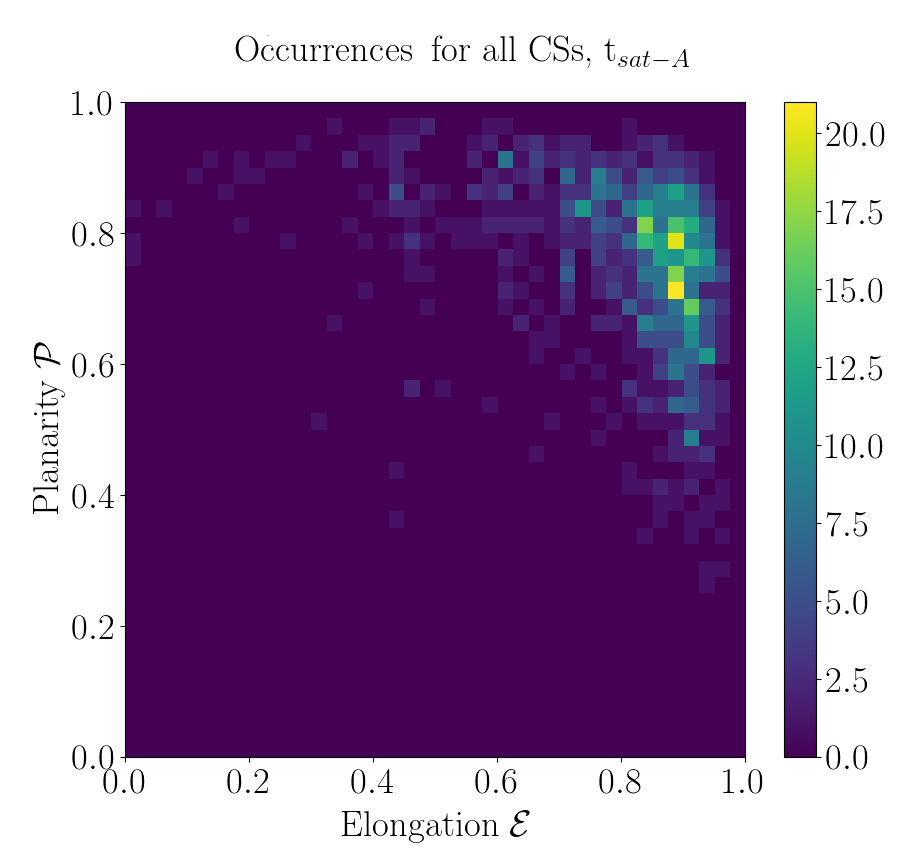}
    \caption{Occurrence distributions of elongation $\mathcal{E}$ versus planarity $\mathcal{P}$ (as computed using the ``non-local'' overall method discussed in Section \ref{sec:definitions}) for SIM A.}
    \label{fig:EP}
\end{figure*}

\begin{figure*}
    \centering
    \includegraphics[width=\textwidth]{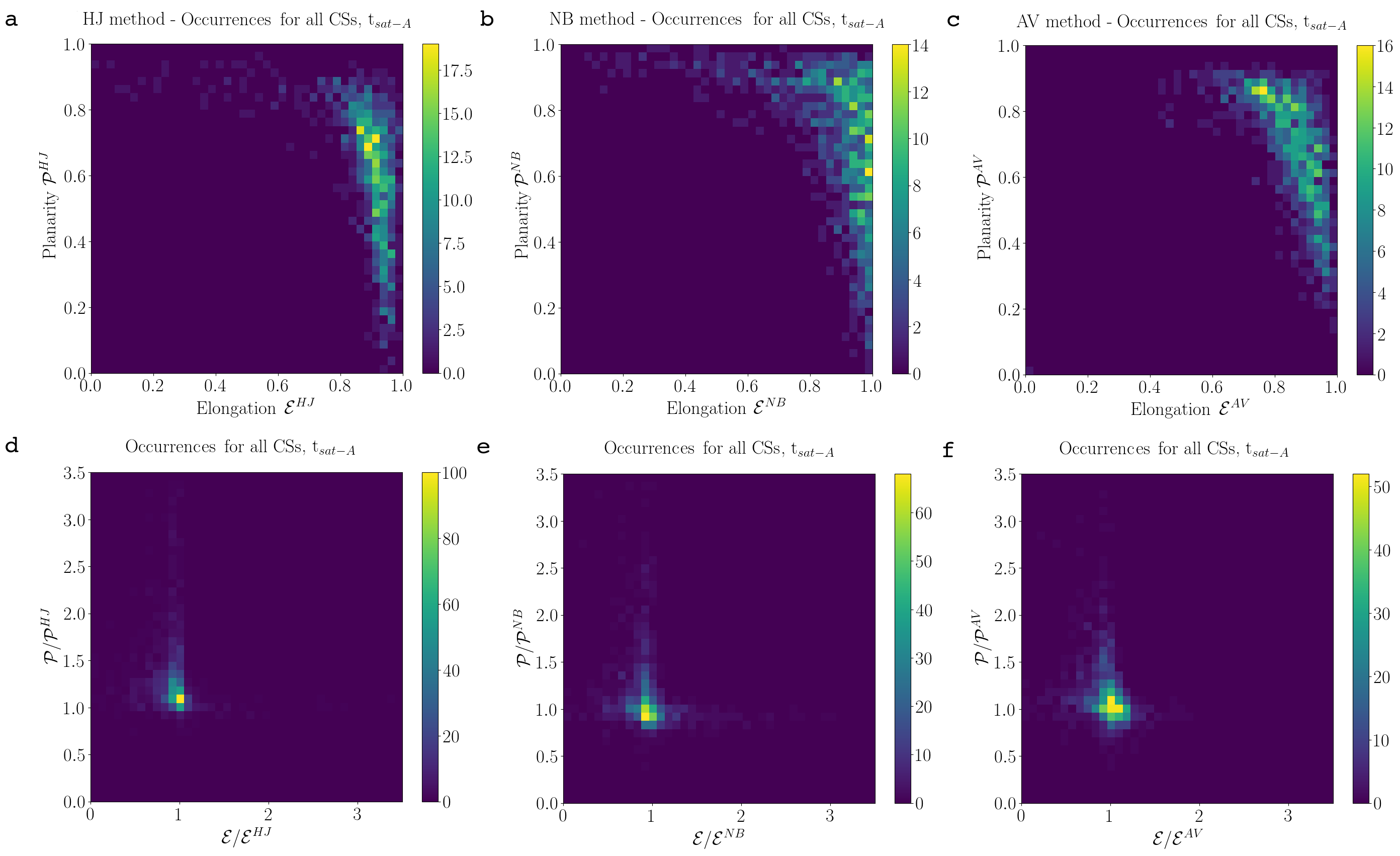}
    \caption{Occurrence distributions of elongation versus planarity, 
    for SIM A. Top row panels: a) HJ method, b) NB method, c) AV method. Bottom row panels: d) $\mathcal{E}/\mathcal{E}^{\met{HJ}}$ vs. $\mathcal{P}/\mathcal{P}^{\met{HJ}}$, e) $\mathcal{E}/\mathcal{E}^{\met{NB}}$ vs. $\mathcal{P}/\mathcal{P}^{\met{NB}}$, f) $\mathcal{E}/\mathcal{E}^{\met{AV}}$ vs. $\mathcal{P}/\mathcal{P}^{\met{AV}}$, thus the comparison between our {\it local} methods and our overall/non-local reference values $\mathcal{E}$ and $\mathcal{P}$. The $x$ and $y$ ranges in panels $d$, $e$ and $f$ have been adapted in order to produce a better visualization. In doing so some cases have been excluded: in panel $d$ 21 over 1043 total current structures (CSs) examined, in panel $e$ 23 over 1043 and in panel $f$ 2 over 1043. }
    \label{fig:cfrEP}
\end{figure*}

In Figure \ref{fig:EP} we show how $\mathcal{E}$ and $\mathcal{P}$ are distributed when defined through $\ell_{max}$, $\ell_{med}$, and $\ell_{min}$ (Equations \ref{eq:el-str} and \ref{eq:pl-str}), while in Figure \ref{fig:cfrEP} we report the results of the estimates from HJ, NB, and AV methods (i.e. distributions for $\mathcal{E}^{\met{HJ}}$ and $\mathcal{P}^{\met{HJ}}$, $\mathcal{E}^{\met{NB}}$ and $\mathcal{P}^{\met{NB}}$, $\mathcal{E}^{\met{AV}}$ and $\mathcal{P}^{\met{AV}}$, respectively, alongside with their ratios to the reference ``non-local'' overall $\mathcal{E}$ and $\mathcal{P}$ values for each structure). 
The physical picture that emerges from Figure \ref{fig:EP} is consistent with highly elongated current structures. Most of the occurrences are concentrated at elongation in the interval $0.8-0.95$ and planarity between $0.5$ and $0.8$ (in particular, for mean, median and standard deviation see first two rows of Table \ref{tab:table-means-methodcomp})
suggesting ``knife blade'' 3D configurations. The results from HJ and AV methods are in very good agreement with the overall occurrence distribution of $\mathcal{E}$ and $\mathcal{P}$, as shown in the top-row panels of Figure \ref{fig:cfrEP}. For what concerns method NB, it gives the same physical picture of the other local/non-local methods, with highly elongated and mostly planar current structures (this is particularly clear from panel (e)), but from panel (b) we also note a tendency to return higher values of planarity and elongation (indeed the distribution is a bit shifted towards the upper-right corner).
The near equivalence of the three methods in evaluating the characteristic shape of the current structures is further confirmed in the bottom row of Figure \ref{fig:cfrEP}. 
More specifically, Figure \ref{fig:cfrEP} (d) shows that the distribution of $\mathcal{E}/\mathcal{E}^{\met{HJ}}$ is peaked around 0.9, and that $\mathcal{P}/\mathcal{P}^{\met{HJ}}$ is peaked around $\sim1$, meaning that for the majority of current structures the true elongation and planarity are almost the same of the ones obtained using HJ method (for mean, median and standard deviation see 3rd and 4th rows Table \ref{tab:table-means-methodcomp}). The same agreement is shown in Figure \ref{fig:cfrEP} (e) between the control values and the NB method estimates, as shown by mean, median and standard deviation reported in Table \ref{tab:table-means-methodcomp} (5th and 6th rows).
Finally, Figure \ref{fig:cfrEP} (f) shows that the AV method performs even better as compared to other local methods and this is confirmed by the mean, median and standard deviation values reported in Table \ref{tab:table-means-methodcomp}, 7th and 8th row. 

Despite the overall picture provides a very good agreement between reference and local methods, we note the presence of a slightly but persistent underestimation of the planarity using the local methods with respect to the reference one, underlined by the mean values reported in Table \ref{tab:table-means-methodcomp}.
Despite this general effect, we deduce that by using local methods, and in particular those employing the {\bf N} tensor, it is possible to correctly estimate the actual shape of current structures. The main advantage of using the {\bf N} tensor is that it can be applied to any point and not necessarily on the center of current structures. Even more importantly, the AV method, which is also the one which performs the best, can be applied with slight modifications to satellite data, once that a proper current threshold is defined.

\begin{table}
\caption{\label{tab:table-means-methodcomp} Mean, median, standard deviation for method comparison}
\centering
\begin{tabular}{cccc}
Quantity & Mean & Median & $\sigma$ \\
      \hline
     $\mathcal{E}$ & 0.8 & 0.9 & 0.4 \\
     $\mathcal{P}$ & 0.7 & 0.8  &  0.2 \\
     $\mathcal{E}/\mathcal{E}^{HJ}$ & 0.9 & 0.9 & 0.2 \\
     $\mathcal{P}/\mathcal{P}^{HJ}$ & 1.3 & 1.2  &  0.4\\
     $\mathcal{E}/\mathcal{E}^{NB}$ & 0.9 & 0.9 & 0.3 \\
     $\mathcal{P}/\mathcal{P}^{NB}$ & 1.1 & 1.0  &  0.4\\
     $\mathcal{E}/\mathcal{E}^{AV}$ & 1.0 & 1.0 & 0.2 \\
     $\mathcal{P}/\mathcal{P}^{AV}$ & 1.1 & 1.0  &  0.3\\
\end{tabular}
\end{table}


\section{Investigating physical characteristics of current structures}\label{sec:results}
 
In this section we investigate the physical features of the structures forming in the three different 3D simulations. 
We will investigate the magnetic configuration shape factors (planarity and elongation), the characteristic eigenvalues (proportional to the characteristic scale lengths) and the orientation of magnetic field and current density. 
We will use the ${\bf N}$ tensor and its eigenvalues, but without restricting our analysis to points owning to the current structures, as for NB and AV methods, thus we will essentially apply the MCA method as described first in \cite{fadanelli_fourspacecraft_2019}. 
The aim is to determine what are the statistical characteristics of magnetic configurations that emerge in our simulations and whether there is any appreciable difference between the configurations found inside and outside current structures. Thus, we will consider separately 1) all points belonging to current structures, 2) the same number of points as in the first case, but belonging to an uniform sampling of the simulation box. These last points are called ``generic''. 
This type of analysis further expands the methodology employed by \cite{fadanelli_fourspacecraft_2019} on satellite data, where, however, the statistical properties of magnetic configurations have been investigated without distinguishing current structures from the rest of the plasma. In particular, the analysis of ``generic points'' can be considered the equivalent of a continuous sampling of satellite data, as done by  \cite{fadanelli_fourspacecraft_2019}. 
We expect that the uniformly picked set would reproduce data collected by satellite along 1D trajectories. Indeed, in the uniform sampling we picked one point every three along each direction. Because the typical dimensions of a current structure are in average big enough to include more than three grid points in at least two directions, for each structure several points are collected, as it is the case for a continuous satellite sampling. Investigating the possible difference between an uniform sampling and synthetic 1D satellite trajectories is out of the scope of the present manuscript.

Furthermore, we compare the above analysis performed on our 3D-3V simulations with an analogous analysis applied to two intervals of high-resolution (burst) data which have been collected as the MMS spacecraft encountered a turbulent magnetosheath region just downstream the bow shock. These same intervals were previously considered by \cite{Stawarz_2019} for a complete analysis of turbulence. In order to be directly comparable with our simulations, a three-step selection on the above-mentioned MMS data has been applied so that the magnetic configuration analysis has been performed only on those data points for which (i) the computation of $\bf{N}$ is precise enough that at least two eigenvalues are well determined, (ii) $\beta\in [0.3,3]$ and (iii) the resolution attained by MMS data is comparable with that of the numerical simulations (after these selections, roughly 20\% of original data are kept; for details, see Appendix \ref{appendix:selection-sat-data}). 


\subsection{Analysis of shape factors for magnetic configurations }\label{subsec:Analysis_shape_factors_magnetic_config}
\begin{figure*}
    \centering
    \includegraphics[width=1.0\textwidth]{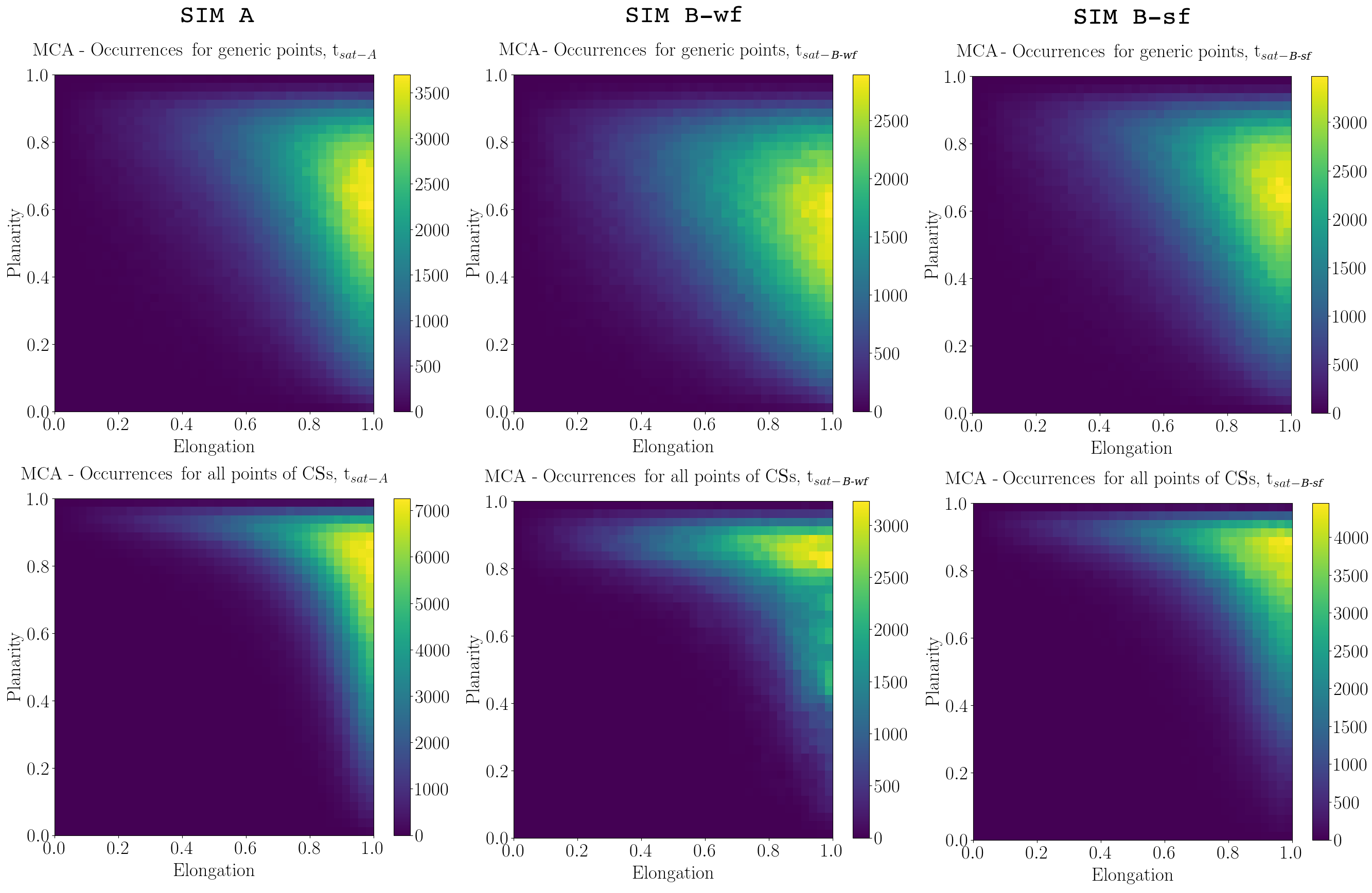}
    \caption{Occurrence distribution of elongation $\mathscr{E}$ versus planarity $\mathscr{P}$ for the three different simulations (three columns), in the top row for generic points, in the bottom row for the points which belong to current structures.} 
    \label{fig:2dhist_plan_and_el}
\end{figure*}

In Figure \ref{fig:2dhist_plan_and_el} we show the occurrence distribution of $\mathscr{E}$ vs $\mathscr{P}$ in our set of 3D simulations, namely SIM A, SIM B-wf, and SIM B-sf (left, center, and right column, respectively): in the top row for generic simulation points (which can be considered as the equivalent of the continuous sampling in the methodology developed by \cite{fadanelli_fourspacecraft_2019}), in the bottom row only for points belonging to what is identified as a ``current structure'' (i.e., those regions where the current density is above $J_{th}$).  

Occurrence distributions which consider only points belonging to current structures (bottom row) or considering generic points (top row) look very different. In particular, if we consider generic points we get a distribution spread in planarity between $\sim 0.2$ and $\sim 0.9$ with a peak around $\sim 0.6-0.7$. Instead, if we consider only points which belong to current structures we get an increase in planarity with the distribution peaking around $\sim 0.8-0.9$ in the $y$-axes. The fact that the distributions of magnetic configurations obtained via generic simulation points does not provide a reliable estimate is even more evident when comparing the results for SIM A presented in Figure \ref{fig:2dhist_plan_and_el} with those obtained in Section \ref{subsec:method_comparison} for the same simulation ({\it i.e.}, see Figures \ref{fig:EP} and \ref{fig:cfrEP}(a-c)). We further point out that for SIM B-wf this discrepancy between the two distributions ({\it viz.}, generic points vs only points belonging to current over-densities) seems to be particularly emphasized. In the following we will give the mean, median and standard deviation for the corresponding 1D-distributions of planarity and elongation.   

\begin{figure*}
    \centering
    \includegraphics[width=\textwidth]{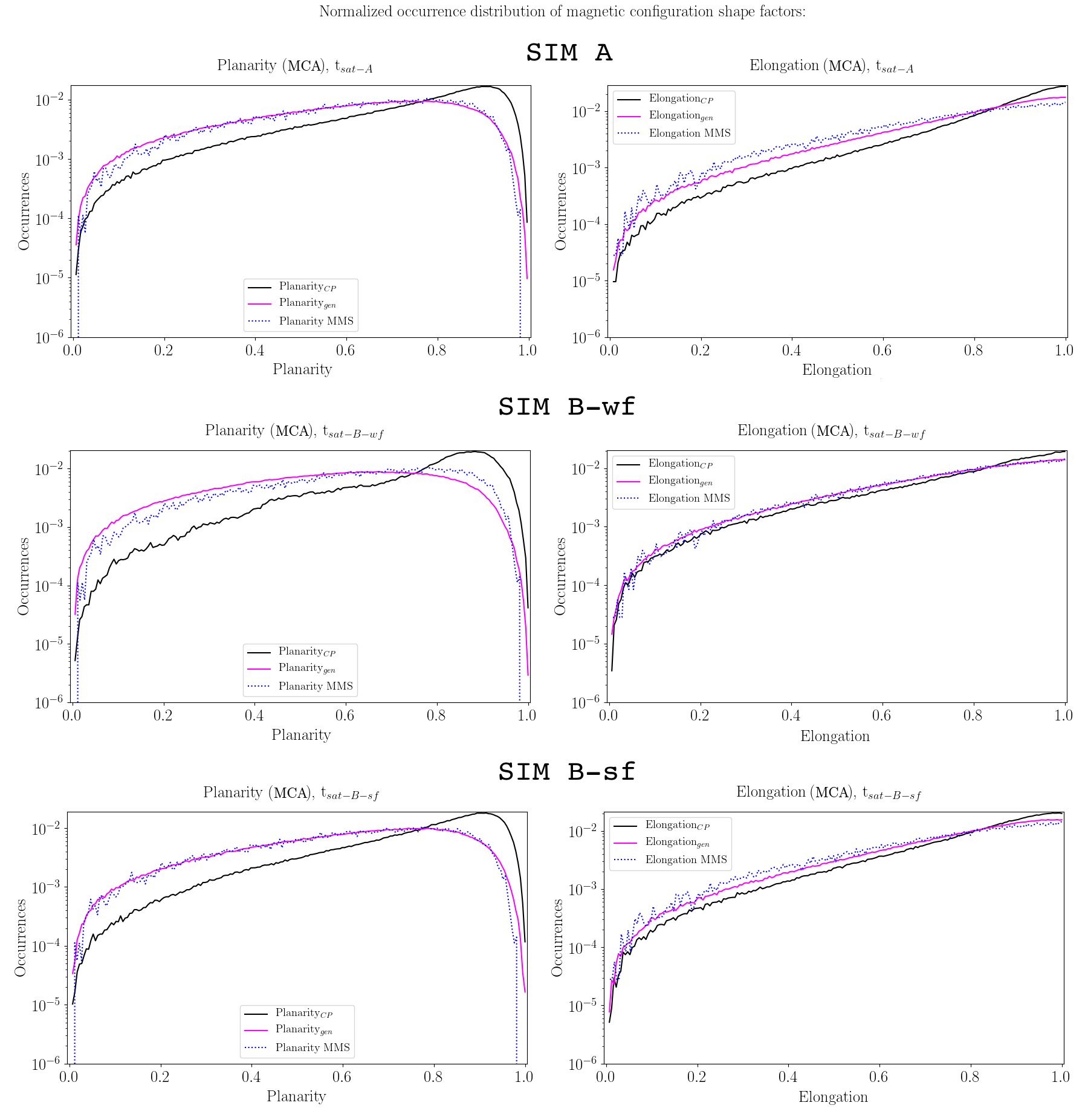}
    \caption{Normalized occurrence distributions of $\mathscr{P}$ and $\mathscr{E}$ for the three simulations SIM A, SIM B-wf and SIM B-sf, distinguishing between statistics on generic points (magenta line) and points which belong to current structures (black line). 
    We superpose to these distributions, obtained from simulations, the ones obtained using satellites data  (dotted blue line) from the two high-resolution magnetosheath intervals analyzed in \cite{Stawarz_2019}, see Appendix \ref{appendix:selection-sat-data} for details.}
    \label{fig:1d_plan_and_el_all}
\end{figure*}

The statistical features of the magnetic configurations can be appreciated in Figure \ref{fig:1d_plan_and_el_all} where we show the 1D normalized occurrence distribution of $\mathscr{E}$ and $\mathscr{P}$ for the three simulations SIM A, SIM B-wf and SIM B-sf for generic points (magenta line) and for points belonging to current structures (black line). We have superposed to these distributions obtained from simulations, the ones obtained using MMS satellites data (dotted blue line) from the two high-resolution magnetosheath intervals analyzed in \cite{Stawarz_2019} (see Appendix \ref{appendix:selection-sat-data} for details). 

As in the color-maps of Figure \ref{fig:2dhist_plan_and_el}, the histograms in Figure \ref{fig:1d_plan_and_el_all} shows an appreciable statistical difference in planarity 
between generic points (magenta lines) and those located inside current structures (black lines). Indeed, if we consider generic points we obtain a $\mathscr{P}$ distributions less skewed towards $\mathscr{P}\approx1$,
with a peak around $0.6-0.8$ (please, note the log-scale for the $y$-axis). 
On the contrary, the normalized histogram of $\mathscr{P}$ for points owning to current structures peaks around $0.8-1.0$. For all three simulations the values of mean, median and standard deviation are reported in Table \ref{tab:table-means-configurations} (1st row for generic points and 2nd row for current structures) and turn out to be the same. 
For the elongation, the 
behavior changes less significantly when considering generic points or current structures. Indeed, the peaks of the occurrence distributions are always close to $1$. Also in this case we report mean, median and standard deviation in Table \ref{tab:table-means-configurations} (3rd row for generic points and 4th row for current structures). Summarizing, the emerging picture indicates a majority of ``blade-like'' magnetic configurations. Commenting on the different behavior between generic points and those located in current structures, this is not surprising since the points located in current structures belong to regions which tend to have a specific shape and a common behavior. In particular, we note that they have an high planarity confirming the intuitive picture of nearly 2D current sheets. On the contrary for generic points, picked up almost from everywhere, also from regions where the current density is low, the magnetic configuration has a different behavior.

\begin{table*}
\caption{\label{tab:table-means-configurations} Mean, median, standard deviation for configuration analysis}
\centering
\begin{tabular}{ccccc}
Quantity & SIM & Mean & Median & 
Standard deviation\\
      \hline
     $\mathscr{P}_{\text{gen}}$ & SIM A, SIM B-wf, SIM B-sf & 0.6 & 0.6  &  0.2 \\
     $\mathscr{P}_{\text{CSs}}$ & SIM A, SIM B-wf, SIM B-sf & 0.7 & 0.8  &  0.2 \\
     $\mathscr{E}_{\text{gen}}$ & SIM A, SIM B-wf, SIM B-sf & 0.8 & 0.8  &  0.2 \\
     $\mathscr{E}_{\text{CSs}}$ & SIM A, SIM B-wf, SIM B-sf & 0.8 & 0.9  &  0.2 \\
     $\log(\sigma_{med,\text{CSs}})$ & SIM B-wf & -2.7 & -2.7 & 0.5 \\
     $\log(\sigma_{med,\text{CSs}})$ & SIM B-sf & -1.7 & -1.7 & 0.5 \\
     $\log(\sigma_{max,\text{CSs}})$ & SIM B-wf & -1.3 & -1.3 & 0.2 \\
     $\log(\sigma_{max,\text{CSs}})$ & SIM B-sf & -0.3 & -0.3 & 0.3 \\
     $\mathscr{E}_{\text{gen,without CSs}}$ & SIM A & 0.8 & 0.8 & 0.2 \\
     $\mathscr{E}_{\text{gen}}$ & SIM A& 0.8 & 0.8 & 0.2\\ 
     $\mathscr{P}_{\text{gen,without CSs}}$ & SIM A & 0.6 & 0.6 & 0.2 \\
     $\mathscr{P}_{\text{gen}}$ & SIM A & 0.6 & 0.6 & 0.2\\
\end{tabular}
\end{table*}

Finally, a major result is the behavior of the distributions for satellite data in agreement with the results of our simulations for generic points. In particular, the agreement is the most evident for SIM B-sf. This agreement is a very good result but not surprising for the following two reasons. First, since there is no selection on the values of $J$ for the analysis performed on MMS time series, it was expected that any agreement with such analysis would have involved the sub-set of ``generic points'' from our simulations. Second, simulation SIM B-sf has a setup which is the 3D equivalent of the one employed in \cite{Califano_2020}, which already proved capable of qualitatively reproducing the turbulent and reconnection regime observed during that period. Thus we expected to find similar features in the configuration shape factors comparing SIM B-sf and these particular MMS intervals.  


\subsection{Analysis of eigenvalues' occurrence distribution and characteristic scale lengths, in simulations and satellite data}

\begin{figure*}
    \centering
    \includegraphics[width=\textwidth]{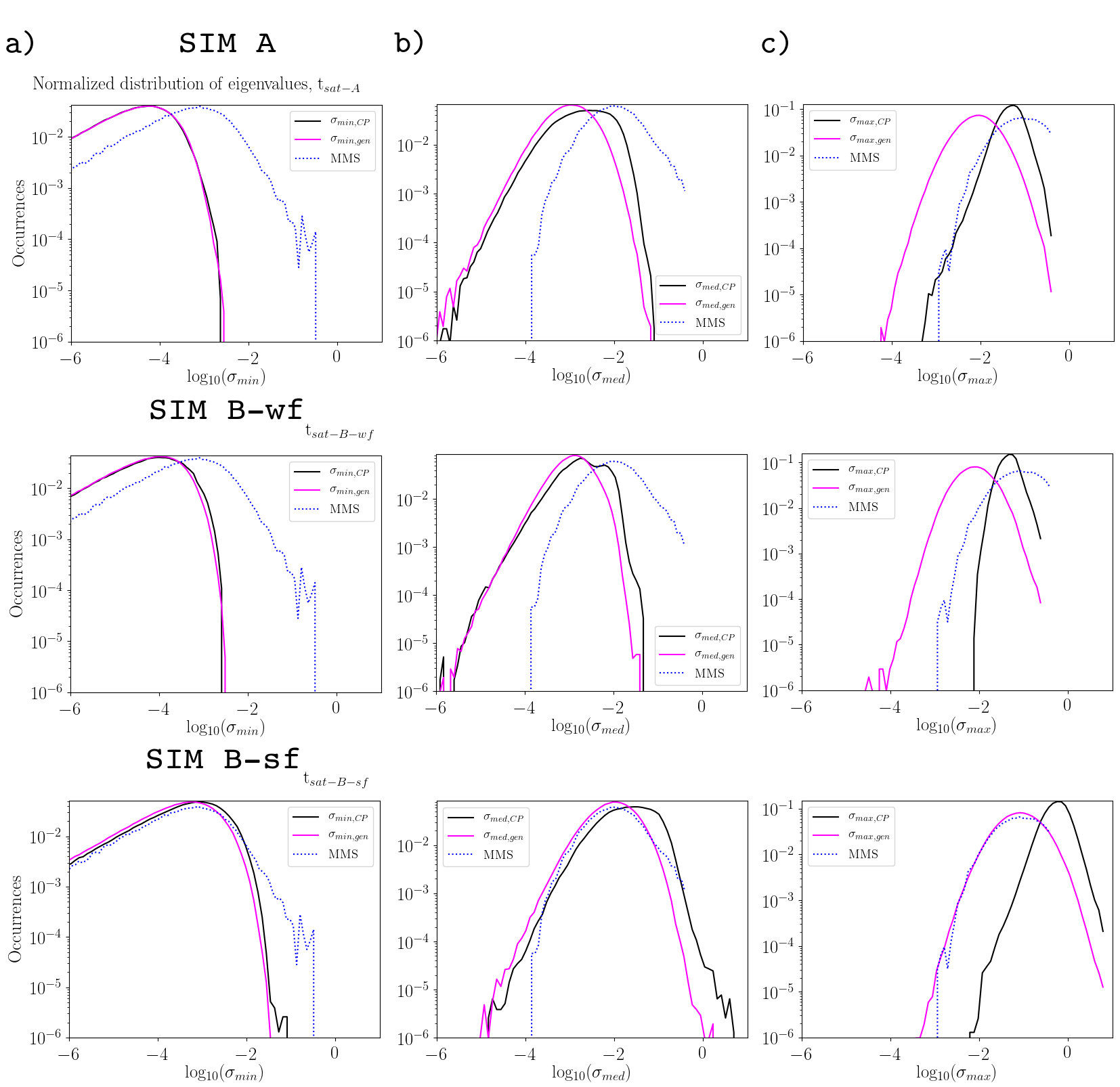}
    \caption{Normalized occurrence distribution of ${\bf N}$'s eigenvalues: $a$) $\sigma_{min}$, $b$) $\sigma_{med}$, $c$) $\sigma_{max}$, for SIM A, SIM B-wf and SIM B-sf, respectively). 
    In magenta, distributions for generic points, in black, for points belonging to current structures. We superposed to these distributions, obtained from simulations, the ones obtained using satellites data (dotted blue line) from the two high-resolution magnetosheath intervals analyzed in \cite{Stawarz_2019}, see Appendix \ref{appendix:selection-sat-data} for details. Note the logarithmic scale on the $x$-axis.}
    \label{fig:1d_eigenvalues_all}
\end{figure*}

In Figure \ref{fig:1d_eigenvalues_all} we show the normalized occurrence distribution of ${\bf N}$'s eigenvalues: $a$) $\sigma_{min}$, $b$) $\sigma_{med}$, $c$) $\sigma_{max}$ for SIM A, SIM B-wf and SIM B-sf.  
We recall that these eigenvalues are interpreted as proportional to the squared inverse of the local variation length (see Section \ref{sec:definitions}). We superpose to these distributions, obtained from simulations, the ones obtained using satellites data (dotted blue line) from the two high-resolution magnetosheath intervals analyzed in \cite{Stawarz_2019} (with the additional selection discussed at the beginning of this Section; see Appendix \ref{appendix:selection-sat-data} for details). In the following, we note that the MMS eigenvalues are normalized using the local value of $d_i$. 

Let us consider first the occurrence distribution of the smallest eigenvalue (which is related to the largest characteristic length of the magnetic configuration). This is shown in column $a)$ of Figure \ref{fig:1d_eigenvalues_all}. For all three simulations there is almost no difference between the distributions for generic points (magenta line) and for points belonging to current peaks (black line) related to this eigenvalue. Indeed the two lines are almost superposed. This means that the typical largest scale length of a ``generic'' region does not differ significantly from that of a structure where the current peaks. Now let us consider instead the occurrence distributions for the other two eigenvalues (related to the median and shortest variation length of the local magnetic configurations; shown in column $b)$ and $c)$ of Figure \ref{fig:1d_eigenvalues_all}, respectively). In all the simulations, the occurrence distributions of these eigenvalues evaluated at generic points significantly differ from the corresponding distributions that are obtained considering only those points belonging to current structures. In particular, the difference is more pronounced for the smallest length scale, i.e. the one associated to the largest eigenvalue. In all cases, when compared to those obtained using generic points, the distributions obtained using only the points belonging to current structures shift towards larger $\sigma$. This means that where the current attains values higher that $J_{th}$, the typical scale length of variation of the magnetic field are smaller (as it is somewhat expected, since the current is the curl of the magnetic field). Comparing SIM B-wf and SIM B-sf, which we remind the reader are identical except for the amplitude of the injected perturbation, one finds that the peak of the occurrence distributions is located at larger values of $\log\sigma_{med}$ and $\log\sigma_{max}$ for simulation SIM B-sf rather than for SIM B-wf. 
Thus, simulation SIM B-sf develops current structures with typical scale length smaller than those of the other simulations. This is true also for generic points, even if less pronounced. 
Such feature is likely a consequence of the fact that the two simulations develop different turbulent regimes (see discussion in Section~\ref{sec:spectral_analysis}). In fact, by injecting larger-amplitude magnetic fluctuations in SIM B-sf than in SIM B-wf, a shallower sub-ion-scale spectrum develops and, consequently, a larger amount of turbulent power reaches the smallest scales (cf. the magnetic-field spectrum in Figure~\ref{fig:spectra}). Whether this is actually due to a weak-like versus strong-like turbulent regime (i.e., if it could be a consequence of whether critical balance and/or dynamic alignment establishes or not) or to other effects (e.g., development of a significant amount of electron-only reconnection events) is beyond the scope of this work and will have to await further investigation, as well as additional simulations.
For current structures, we report the mean, median, standard deviation values in Table \ref{tab:table-means-configurations}, 5rd-8th rows, comparing quantitatively SIM B-wf and SIM B-sf.

Concerning the comparison with the distributions extracted from MMS data \cite{Stawarz_2019}, 
we note a very good agreement with the distributions for generic points for SIM B-st. As for the previous Section \ref{subsec:Analysis_shape_factors_magnetic_config}, this very good agreement with SIM B-sf was somewhat expected since a similar simulation setup in the 2D case already proved to be capable to qualitatively reproducing the turbulent and reconnection regime observed in these intervals \cite{Califano_2020}.
Thus, we proved once again that a relatively large-amplitude injection close to the ion-kinetic scale is able to reproduce several features that are observed by MMS in the turbulent magnetosheath past the bow shock--in particular, the development of similar 3D structures. Although such setup constitutes a quite interesting hint, the actual mechanism that could be behind such peculiar injection (e.g., the bow shock itself or the subsequent occurrence of micro-instabilities) is still unclear and requires further investigations, especially from the observational point of view.


\subsection{Analysis of the orientation of magnetic field and current density}

We analyze the orientation of the direction ${\bf e}_{min}$ along which we measure the smallest eigenvalue with respect to the direction of the local magnetic field and current density by computing $|{\bf b}\cdot{\bf e}_{min}|$ and $|{\bf j}\cdot{\bf e}_{min}|$, respectively. Here ${\bf b}$ and ${\bf j}$ are the unit vectors of the local magnetic field and current density. As it was done in previous Sections, this calculation is performed both on generic simulation points and also restricting only to points which belong to current structures. 

\begin{figure*}
    \centering
    \includegraphics[width=\textwidth]{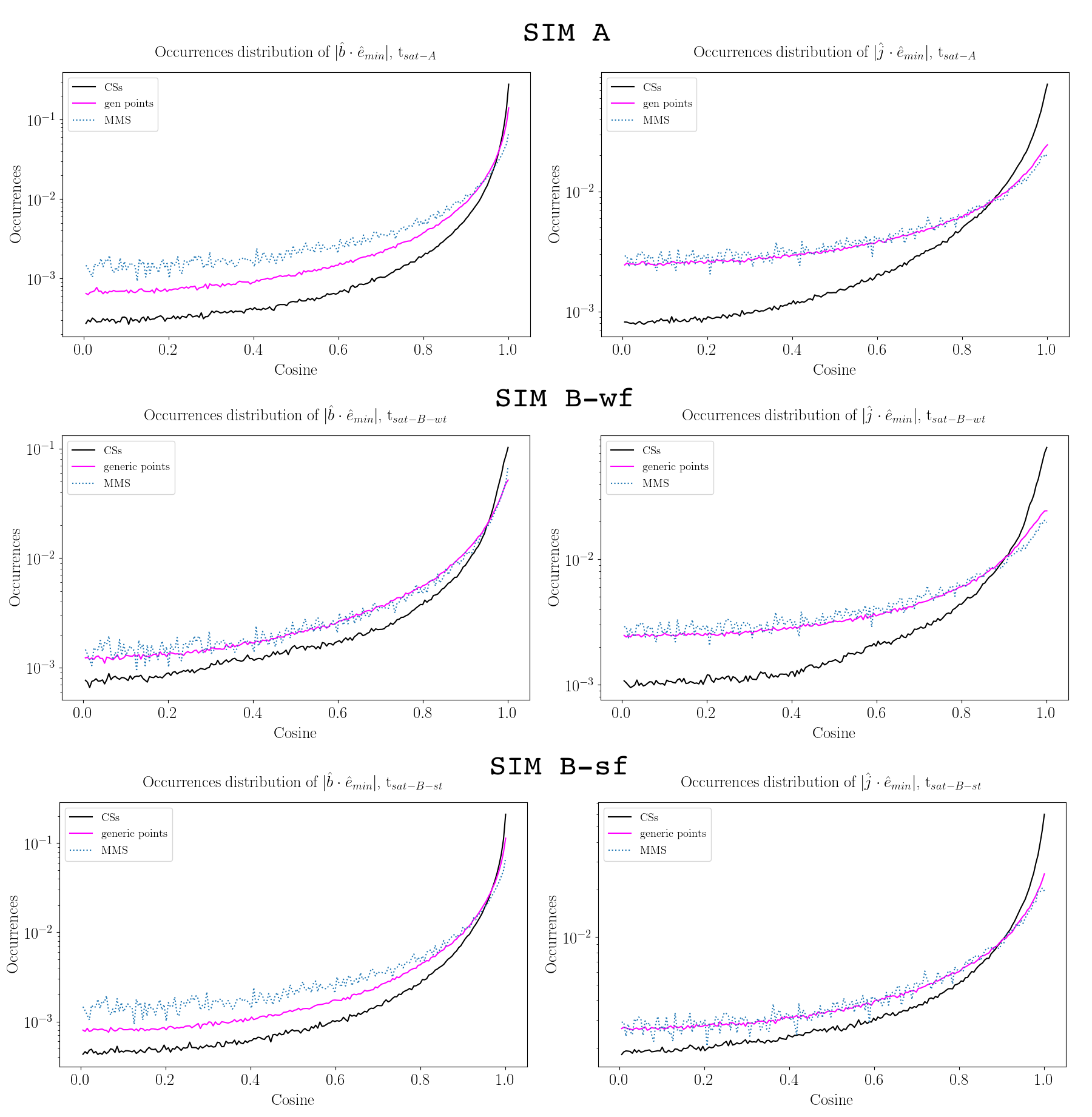}
    \caption{normalized occurrence distribution of $|{\bf b}\cdot{\bf e}_{min}|$ (left column) and $|{\bf j}\cdot{\bf e}_{min}|$ (right column) for all three simulations, in magenta the distribution for generic points, in black that for points belonging to current structures. The distribution refers to times at fully-developed turbulence, in particular $t_{sat_A}$ for SIM A, $t_{sat_B-wf}$ for SIM B-wf and $t_{sat_B-sf}$ for SIM B-sf. The distribution refers to times at fully-developed turbulence, in particular $t_{sat_A}$ for SIM A, $t_{sat_B-wf}$ for SIM B-wf and $t_{sat_B-sf}$ for SIM B-sf. We superposed to these distributions, obtained from simulations, the ones obtained using satellites data (dotted blue line) from the two high-resolution magnetosheath intervals analyzed in \cite{Stawarz_2019}, see Appendix \ref{appendix:selection-sat-data} for details.}
    \label{fig:alignment_all}
\end{figure*}

In Figure \ref{fig:alignment_all} we show the normalized occurrence distribution of $|{\bf b}\cdot{\bf e}_{min}|$ (left column) and $|{\bf j}\cdot{\bf e}_{min}|$ (right column) for all three simulations for generic points (magenta) and for points belonging to current structures (black). 
We superpose to these distributions those obtained by using satellites data (dotted blue line) from the two high-resolution magnetosheath intervals analyzed in \cite{Stawarz_2019} (see Appendix \ref{appendix:selection-sat-data} for details).

Both the local magnetic field and ${\bf j}$ are well aligned to ${\bf e}_{min}$ for all three simulations. In particular, we note that the alignment is stronger for points which belong to current peaks, which means that for these configurations the alignment between magnetic field and ${\bf e}_{min}$ (and the same for ${\bf j}$) is the best one. The strong alignment between ${\bf j}$ and ${\bf e}_{min}$ is expected since by definition the derivatives of ${\bf B}$ perpendicular to ${\bf e}_{min}$ are the strongest ones. Instead, the alignment between the magnetic field and ${\bf e}_{min}$ is less obvious and could depend on the specific environment that is being considered and/or on the initial parameters of our simulations. In general, we expect that the local magnetic field and the current density tend to align only when there is a significant component of ``guide field'' in the structure (opposed to those configurations where $B$ vanishes within the current structure, as, for instance, in the typical setup that is employed to study magnetic reconnection without a guide field). Moreover, the good alignment between $e_{min}$ and B could also be due to Beltramization of the flow, namely the alignment between current and magnetic field, which is typical of small-scale turbulent structures (see e.g. \cite{degiorgio-2017-beltramization}).

Also in this case, the comparison with the observative data from MMS \cite{Stawarz_2019} is very good. In particular, we note that for the alignment $|{\bf b}\cdot{\bf e}_{min}|$ the best agreement between MMS data and simulations ``generic points'' is found with SIM B-wf rather than with SIM B-sf.
This is probably due to the fact that $|{\bf b}\cdot{\bf e}_{min}|$ is strongly affected by the presence of a guide field within the current structure. In fact, if $\delta B/B_0$ is small enough as it is for SIM B-wf, the guide field is less distorted by the turbulent dynamics, and also the direction of weakest variation of the emerging current structures will align better with ${\bf b}$. On the other hand, when large $\delta B/B_0$ fluctuations are injected, there is a significant distortion of the background magnetic field. As a result, the emerging structures can be embedded in magnetic shears where, with respect to the plane perpendicular to ${\bf e}_{min}$, there is a weak (or vanishing) guide field. 

We note, as in the discussion of Figures \ref{fig:1d_plan_and_el_all} and \ref{fig:1d_eigenvalues_all}, that the agreement of the MMS distribution with the ``generic points'' rather than with the dataset of points belonging to current structures was to be expected. Indeed, there is no selection on the values of $J$ for the analysis performed on MMS time series. 

\subsection{Generic points or ``background''}\label{sec:background}

\begin{figure*}
    \centering
    \includegraphics[width=\textwidth]{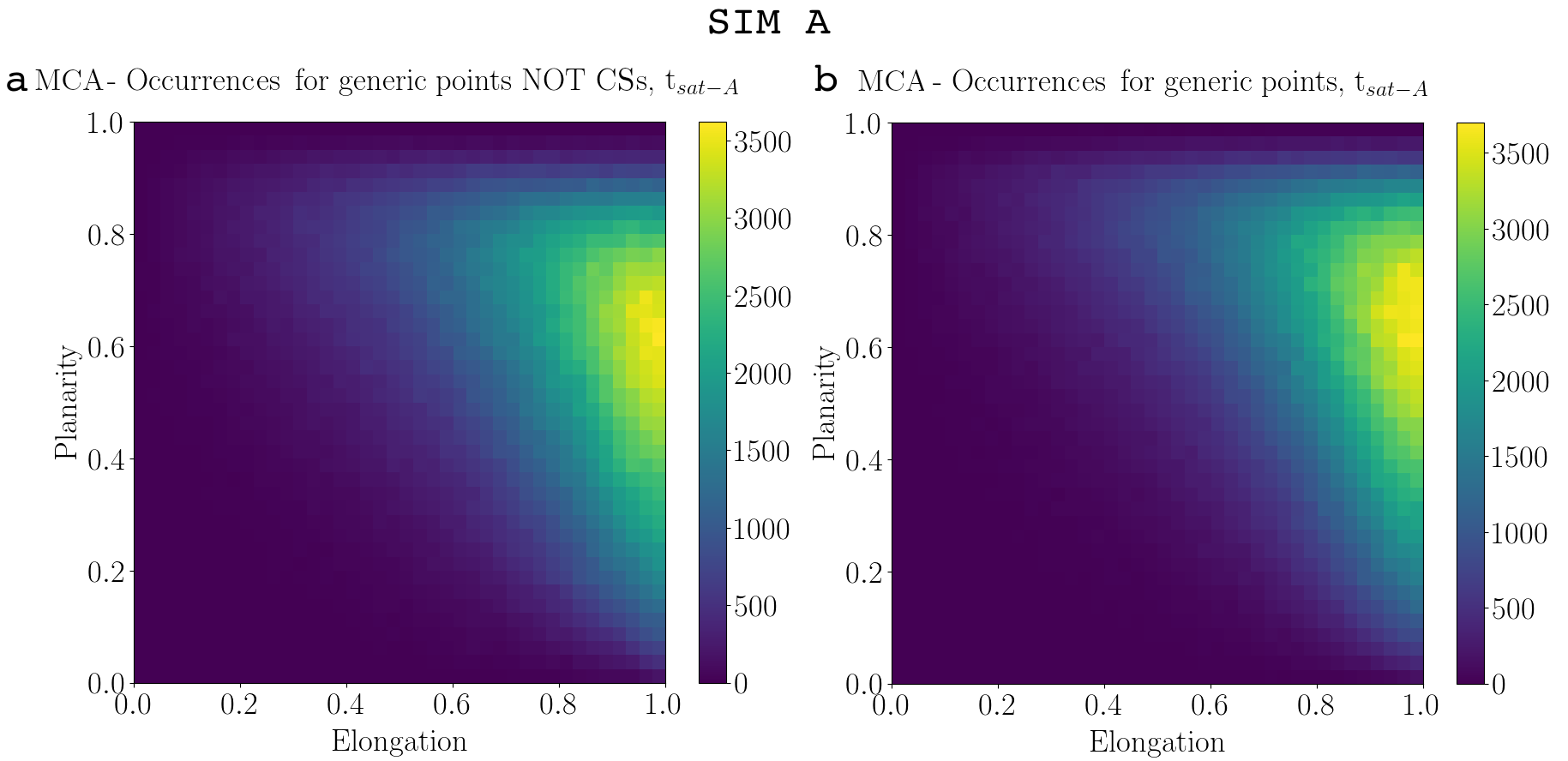}
    \caption{For $t_{sat_A}$ SIM A, occurrence distribution of elongation versus planarity for a) ``generic'' points excluding the ones which belong to current structures (roughly $4\%$), b) ``generic'' points.}
    \label{fig:confronto_background}
\end{figure*}

We briefly explain what implies to perform a statistical analysis on ``generic'' points of simulations or, equivalently, on long continuous sampling in observation data, referring to the methodology proposed in \cite{fadanelli_fourspacecraft_2019}. First of all, we want to make sure that the ensemble of ``generic'' points picked in our analysis constitute a somewhat statistically representative sub-group of the whole grid points of the simulation, while simultaneously being comparable to the total number of points that belong to the current structures in such simulation (which is instead the sub-set used when restricting the analysis to these structures). Let us consider, for instance, SIM A, in which the total number of grid points is $352*352*198 \sim 2.5\times10^7$. 
In this case, the sub-set of ``generic'' points has been created by considering one point every 27 (corresponding to a collection of $\sim10^6$ points), i.e. around $\sim 4\%$ of the original set. Analogously, the total number of points which overcome the threshold on current density at time $t_{sat_A}$ are $\sim1.1\times10^6$ (i.e., roughly $4\%$ of the total). Moreover, in our sub-set of ``generic'' points those which overcome the threshold on current density are $4.8\times10^4$, again roughly $4\%$ of the sub-set. Thus, the sub-set of ``generic'' points that is being considered should adequately represent the ensemble of points of our simulation box. 
As we just discussed, the points which constitute the current structures are always a small percentage of the total number of points in a set or sub-set (i.e., the filling factor of the current structures in a volume is usually very small). Therefore, a natural question to ask is the following: can the behavior of a such small sub-group emerge when we consider histograms and plots which refer to ``generic'' points? (or, equivalently, when we consider a long continuous sample in observation data?) Based on our analysis, the answer to this question is \textit{no}. This can be seen in Figure \ref{fig:confronto_background}, which shows the occurrence distribution of elongation $\mathscr{E}$ versus planarity $\mathscr{P}$ for a) only those ``generic'' points that do not belong to current structures (roughly $4\%$), b) the whole sub-set of ``generic'' points (i.e., including those belonging to current structures). In fact, there is no noticeable difference between the two distributions. In particular, for both cases the elongation and planarity have the same mean, median and standard deviation (see Table~\ref{tab:table-means-configurations}, rows 9 to 12).

The behavior of the current structures cannot emerge if we consider “generic” points, i.e. the equivalent of a long continuous sample in satellite data without introducing any additional selection on the turbulent time series. Therefore, in order to systematically study the statistical behavior of the shape of current structures in satellites data, we suggest to fix a threshold on the current density and to perform the analysis only on those regions that overcome this threshold. This procedure is similar to what has been done when applying the AV method discussed in Section \ref{sec:shape-current-structures} to our numerical simulations. A similar kind of suggestion (to use a threshold to identify strong current points) has been proposed also in other works such as \cite{BRUNO20011201,Greco-2009-PhysRevE.80.046401,Osman-2012-PhysRevLett.108.261103,sorriso-2018}.

\section{Conclusions}\label{sec:conclusions}

In this work we have conducted a wide-spanning analysis of overall/non-local current structures and local magnetic configurations emerging in plasma turbulence. The analysis is based on three 3D-3V HVM simulations with different box sizes, resolution and energy injection scales.
We focused first on the characterization of the shape of current structures. Our ``reference method'' defines two parameters, elongation $\mathcal{E}$ and planarity $\mathcal{P}$, which can be calculated in a simulation from the three characteristic dimensions of any structure ($\ell_{max}$, $\ell_{med}$ and $\ell_{min}$). We have shown how it is possible to reliably estimate the shape of a current structure also by employing three different \textit{local} methods, namely HJ, NB and AV (see Section~\ref{sec:shape-current-structures} for details on the methods). The rigorous computation of $\mathcal{E}$ and $\mathcal{P}$ via the ``reference'' non-local method can be performed only on simulation data. The HJ and NB methods too can be performed only on simulation data, since they require to be applied on the center of a current structure (whose precise position can be known only on simulation data). These two ``local'' methods are faster than the other two (i.e., the ``non-local'' and the AV methods), since they require to be computed on a restricted number of points (namely, only on the local maxima of the current density). Finally, the AV method, the one in best agreement with the results of the ``reference'' method, can be applied with minor modifications also on spacecraft data, once a threshold on the current density has been properly defined on the time series under consideration. 
Based on these four methods (i.e., ``reference/non-local'', HJ, NB and AV), we have analyzed the distribution of shape factors (i.e., planarity and elongation) for the emerging current structures, with the result that all methods coherently find that they are composed of mainly ``knife-blade''-like structures. This picture is different from the one that emerges in \cite{meyrand_2013} where they claim the presence of mostly cigar-like structures (they use the expression “filament like”) through visual inspection of their 3D Hall-MHD simulations of turbulence. This suggests that ion-kinetic effects (i.e., beyond just the Hall term) and/or electron-inertia terms could significantly affect (and likely be required to correctly describe) the development of current structure in plasma turbulence across the transition range and at sub-ion scales.

Additionally, we studied the local magnetic configurations by performing an analysis on the simulation data similar to the one proposed by \cite{fadanelli_fourspacecraft_2019} for satellite data. In particular, we have investigated the magnetic configurations by analyzing the distribution of their planarity and elongation, of their three characteristic scale lengths, and of their orientation with respect to the magnetic-field and current-density directions. Such analysis has been performed both on all points belonging to current structures as well as for ``generic'' points belonging to a uniform sampling of the simulation box (the aim of this latter set being to mimic long and continuous time series from satellite data). In general, we found different results when we apply the analysis only to those points belonging to current structures or to ``generic'' points in the simulation domain (i.e., including, but not limited to, structures).
In particular, the main statistical difference between generic points and those located inside current structure is in the results obtained for the distribution of planarity $\mathscr{P}$:  ``knife-blade'' shapes are more likely present when considering only those points where the current density is above a certain threshold $J_{th}$, while the abundance of ``thicker'' sheets (or ``ellipsoids'') is enhanced when points below $J_{th}$ are included. 
The behavior of the distributions for generic points for planarity $\mathscr{P}$ and elongation $\mathscr{E}$ is coherent with the distribution we obtain if we apply the same kind of analysis to high-resolution (burst) MMS data collected from the two turbulence crossings in the magnetosheath and analyzed previously by \cite{Stawarz_2019}, and selected in order to fit simulations' characteristics (see Appendix \ref{appendix:selection-sat-data} for details). 

In the analysis of variation scale lengths, we found a sensible difference in the distributions for the largest eigenvalue between generic points and points within structures, suggesting that the smallest characteristic length-scale $\ell_{min}$ is significantly shorter for current structures.
We noted a difference in this context also between SIM B-wf and SIM B-sf, for which all the three characteristic length scales are shorter in the strong-forcing scenario (SIM B-sf) than they are when lower-amplitude fluctuations are injected (SIM B-wf). We remind the reader that in both simulations SIM B-wf and SIM B-sf the energy is injected only slightly above the ion characteristic scales, as done in \cite{Califano_2020}, that was able to reproduce, in a simplified 2D-3V configuration, the electron-only turbulent regime observed by MMS past the bow shock \cite{phan_electron_2018,Stawarz_2019}.
Concerning the comparison with the distributions extracted from these precise MMS data \cite{phan_electron_2018,Stawarz_2019}, we found a very good agreement only with the distributions for generic points from SIM B-sf, thus suggesting that similar, small-scale current structures require a high level of forcing acting close to the ion-kinetic scales.

Finally, from the analysis of the local orientation of the magnetic field and the current density with the minimum variance direction $e_{min}$, we found strong alignments for both fields, more pronounced for points which belong to current structures. Also in this case, the comparison with the observative data from MMS \cite{phan_electron_2018,Stawarz_2019} is very good.  In particular, we note a good agreement with the distributions for generic points in SIM B-wf.
 
All the analysis presented in this work clearly highlights how magnetic configurations inside current structure exhibit peculiar features that can be retrieved only by considering solely those points where $J$ attains values above a certain threshold.
Indeed, we have shown in Section \ref{sec:background} that the behavior of the current structures cannot emerge if we consider “generic” points, i.e., analogous to consider a long continuous time series in satellite data without any further selection based on the values of $J$, since for such a sample the number of points belonging to current structures will be only a small percentage of the total. Therefore, in order to systematically study the shape of current structures in satellites data, we suggest to fix a proper threshold on the current density and consequently consider only regions that overcome this threshold for the subsequent analysis. Even better, one could apply the AV method that we have described in Section \ref{sec:shape-current-structures} in order to isolate different structures (after applying straightforward modifications in order to adapt the method to simple 1D time series rather than to complex 3D spatial domains).

In conclusion, the results reported in this paper would be useful not only for the analysis of turbulence simulations, but also for observative studies. Indeed, 1) we have validated the possibility to apply local methods, which are the only ones applicable on satellite data, to infer the overall/non-local shape of current structures; 2) we conjecture that imposing a proper threshold on the current density would benefit for the statistically study of current structures in satellite data; 
3) we have provided an overview of local magnetic configurations emerging in different turbulence regimes, also stressing the different behavior that is found when considering exclusively points within current structures with respect to what emerges from the points belonging to the rest of the turbulent environment; 4) 
we have shown via such magnetic configuration analysis that, when there is a mechanism that inject relatively high-level fluctuations close to the ion-kinetic scales, our 3D-3V simulation can reproduce the structures that emerge in MMS data for the periods studied by \cite{phan_electron_2018,Stawarz_2019}. \\

\noindent {\bf Acknowledgements}
This work has received funding from the European Unions
Horizon 2020 research and innovation programme under grant agreement No. 776262 (AIDA, www.aida-space.eu). We acknowledge the CINECA ISCRA initiative and the EU PRACE initiative (grant n.~2017174107) for awarding us access to the supercomputer Marconi, CINECA, Italy, where the calculations were performed. FC thank Dr. M. Guarrasi (CINECA, Italy) for his essential contribution for his contribution on code implementation on Marconi. S.S.C. was supported by NASA Grant No.~NNX16AK09G issued through the Heliophysics Supporting Research Program, and by the Max-Planck/Princeton Center for Plasma Physics (NSF grant PHY-1804048).

\appendix
\section{Alternative definitions for structure's dimensions} \label{appendix:alternative_dimensions}

\begin{figure*}
    \centering
    \includegraphics[width=\textwidth]{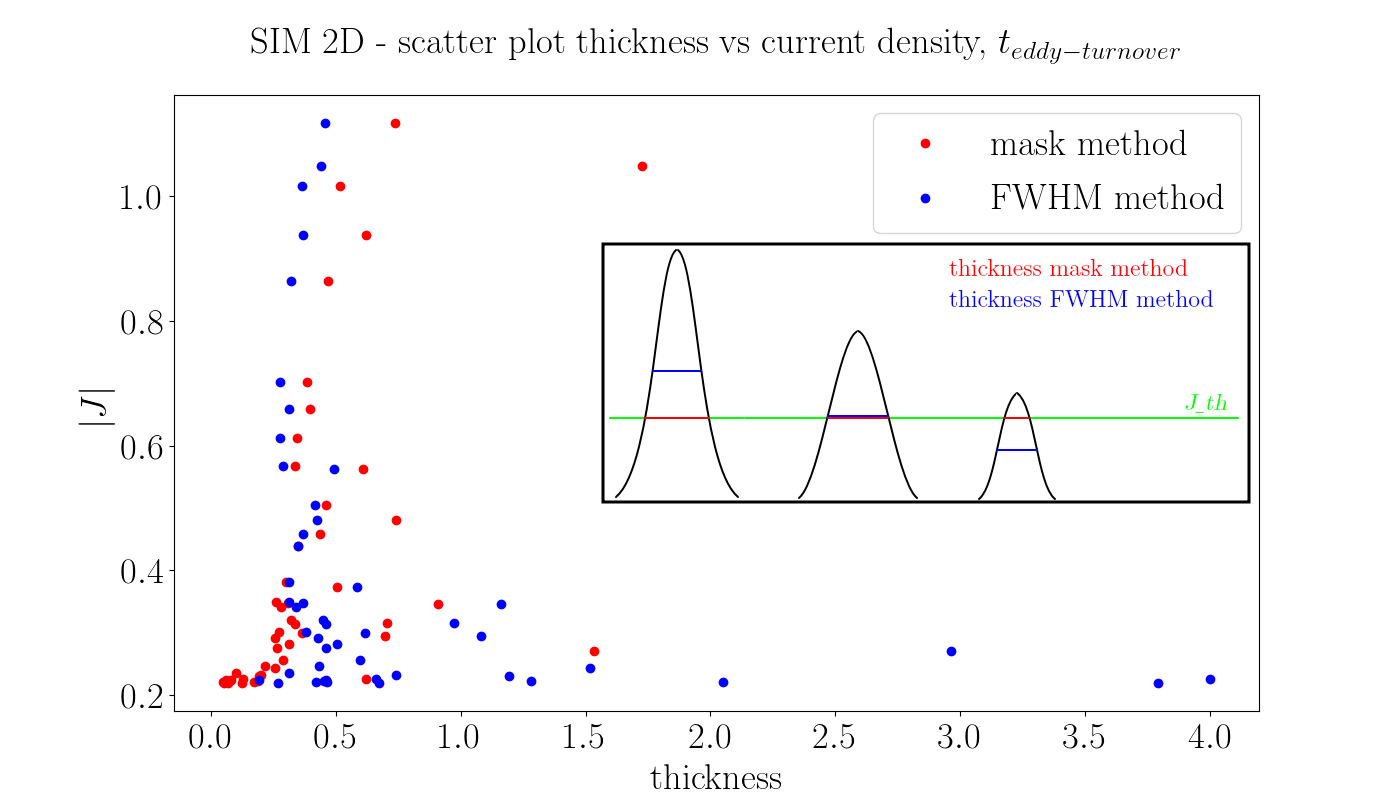}
    \caption{scatter plot thickness vs current density for SIM 2D at $t_{eddy-turnover_{2D}}$, to show the differences between the two methods in estimating thickness. The current density is computed at the center of each current structure.  With ``mask method'' we mean the method to find thickness explained in Section \ref{sec:definitions}, which uses the maximum distance between points satisfying $J\ge J_{th}$. On the contrary, with ``FWHM method'' we refer to the alternative method explained in Appendix \ref{appendix:alternative_dimensions} which computes thickness using full width at half maximum of the interpolated profile of $J$.  In the sub-panel we show a schematic representation to explain the difference in estimations between the "FWHM method" and the "mask method". }
    \label{fig:appendix:scatter:thickness:current:scheme}
\end{figure*}

\begin{figure*}
    \centering
    \includegraphics[width=0.6\textwidth]{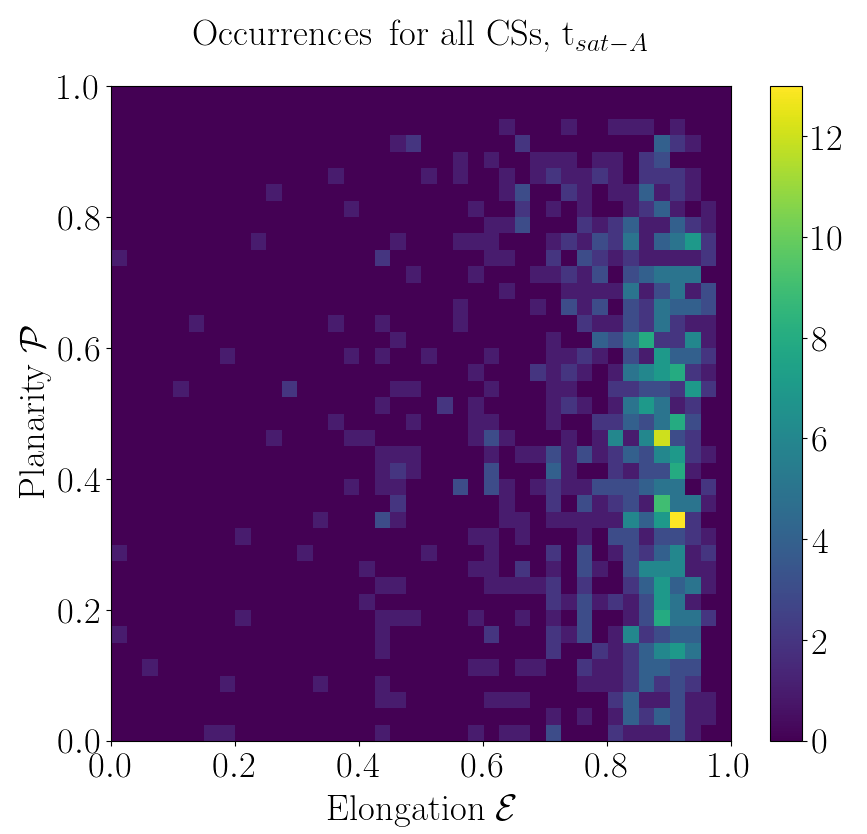}
    \caption{Occurrence distributions of elongation $\mathcal{E}$ versus planarity $\mathcal{P}$, at $t=t_{sat_A}$ for SIM A. Planarity here is computed using the thickness given by the alternative method explaneid into Appendix \ref{appendix:alternative_dimensions}. }
    \label{fig:appendix:EP}
\end{figure*}

In our work, we defined the thickness $\ell_{min}$ as the maximum distance between two points with $J\ge J_{th}$ along the direction of strongest variation of the current density, as given by the eigenvectors of the Hessian matrix. We present here an alternative definition for thickness, as proposed by \cite{zhdankin_statistical_2013}, and we show how some of our results change when using this different definition. In particular, we can alternatively define thickness considering the interpolated profile of $J$ along the direction of strongest variation given by the Hessian matrix and computing it as the width at half maximum of this profile. 
The two methods can produce different estimations for thickness depending on the value of the current density peak. In Figure \ref{fig:appendix:scatter:thickness:current:scheme} we show the scatter plot thickness versus current density for SIM 2D at $t_{eddy-turnover_{2D}}$: in red the estimations given by the method used in the main text (maximum distance between two points with $J\ge J_{th}$, for brevity, called ``mask method''), while in blue the ones given by the method used in this appendix (full width at half maximum of the current density profile, called ``FWHM method'').  We chose the 2D simulations for this comparison and the eddy-turnover time rather than the saturation time just because in this simulation, at this time, current structures are not so many to make the scatter plot difficult to be interpreted.  The current density is computed at the center of each current structure considered. We can see that the two estimations turn out to be different, especially at low values of current density.  In the sub-panel, we show a schematic representation to explain the difference in estimations between the ``FWHM method'' and the ``mask method''. In particular, when the current density at the center of the specific structure is greater than $2J_{th}$, thickness estimated using the ``FWHM'' method is generally bigger than the one estimated using ``mask method'', on the contrary when the current density is lesser than $2J_{th}$ the thickness obtained with the ``mask method'' is the biggest one. The two values are almost equal when $J\sim 2J_{th}$. 
If we use this alternative definition of thickness for computing the structure's shape factor planarity, some features change (we note that the estimation of elongation does not change since we haven't changed the definitions of width and length). 
In particular, in Figure \ref{fig:appendix:EP} we show the new distribution for $\mathcal{E}$ and $\mathcal{P}$ which emerges, which is a bit different from the one of Figure \ref{fig:EP}. In particular, to quantify, now we have: 
for the elongation mean$\sim0.8$, median $\sim 0.9$, standard deviation $\sim 0.2$, instead for planarity mean $\sim0.4$, median $\sim 0.4$ and standard deviation $\sim 0.2$. The distribution is lesser in agreement with the ones for HJ, NB, and AV method showed in Figure \ref{fig:cfrEP}, and the most of the structures have a planarity in between 0.2-0.6, thus are more filament-like. 

The width (the intermediate length) is defined as the maximum distance between points having $J>J_{th}$, in the plane perpendicular to $e_{min}$ and passing from the current peak of the structure. For the sake of coherence, we prefer the ''mask method'' for defining the thickness. In such a way, a cylindrical structure would have, as it is, coherent values for both width and thickness, independently from the intensity of the current.

\section{Selection and processing of MMS data}\label{appendix:selection-sat-data}
In a number of passages in the main text, we referred to the two  burst-resolution MMS data intervals analyzed in \cite{Stawarz_2019}, which have been chosen here for comparison with the results of numerical experiments detailed in Section \ref{sec:simulation}. Choice of these MMS data intervals is motivated by the fact that our simulation set-up (in particular for SIM B-wt and SIM B-st) is similar to that used in \cite{Califano_2020}, which was able to reproduce the turbulent and reconnection regime observed during that period. 

Throughout our analyses, MMS data relative to the magnetic field has been taken from the FluxGate Magnetometer (FGM - see \cite{russell_magnetic_2015}), while the Fast Plasma Investigation (FPI - \cite{pollock_fast_2016}) has provided the density and pressure measures which contributed to determine the ion plasma beta and ion inertial lengths. All these aforementioned fields have been interpolated onto the MMS2 magnetic field data, and averaged over the four spacecraft fleet. The $\bf{J}$ value has been obtained by applying the curlometer technique by \cite{dunlop_analysis_1988} and the $\bf{N}$ tensor was derived as in \cite{fadanelli_fourspacecraft_2019} i.e. by performing a linear estimation of the magnetic field gradient, then combining the resulting values with the four-spacecraft averaged magnetic field data.

In order to compare results from MCA applied over simulation and MMS data, we need apply three different levels of selection to the latter. In particular: 1) we select points for which at least two of the eigenvalues of the $\bf{N}$ matrix are well-determined by MMS measurements; 2) we select data with $\beta\in[0.3,3]$; 3) we select data with a resolution comparable with the one of our simulations. 

More in detail, the first selection has been performed exactly like in \cite{fadanelli_fourspacecraft_2019} i.e. by calculating the average inter-spacecraft distance $\ell_{SC}$ at each instant and then setting at $(\frac{\delta B}{\ell_{SC} B})^2$ a minimal resolution threshold for the eigenvalues of $\bf{N}$, being understood that any eigenvalue below such a threshold is not well-resolved. Requiring that at least two of the eigenvalues are well-resolved signifies that, on one hand, we accept uncertainty on the least eigenvalue and elongation measures but, on the other hand, we are still able to determine correctly the direction of minimum variation (this is because the corresponding eigenvector is, by construction, perpendicular to the other two, which are well determined). The requirement that two eigenvalues are well resolved is generally a good compromise between the need for precision and the difficulty of determining correctly the smallest eigenvalue, which is generally extremely small and therefore easily falls below the MMS resolution threshold. Choosing the two-well-determined-eigenvalues selection criterion for the two turbulence intervals we consider here implies that over the $99 \%$ of original data is retained by this first procedure. 

The second selection just described is performed over the ion plasma beta obtained by the ratio of spacecraft-averaged kinetic and magnetic pressures at each datapoint. For the turbulence intervals we consider in this work, this is the procedure which leads to the largest reduction in the availabel dataset, which gets reduced to about $20 \%$ of its original size through it. 

With the third and last filtering of MMS data we intend to eliminate all those points for which simulation and satellite data have different resolutions. 
To compare precision of MMS data and of simulations, we introduce two quantities that we call ``resolution factors'' and are defined as follows:
\begin{itemize}
\item  for the MMS (satellite) measurements, the resolution factor is the inter-spacecraft separation divided by ion inertial length (we remember that the only possible derivative calculation in this case follows from a linear interpolation of satellite measures)
\item  for the simulations, the resolution factor is $max(\frac{3 dx}{d_i},\frac{3 dy}{d_i},\frac{3 dz}{d_i})$ and by this definition we intend to acknowledge that the effective minimal distance over which a HVM numerical simulation can well represent plasma physics should be several times (here we have chosen three) the distance between neighboring data points. 
\end{itemize}
When considering MMS data analyzed in \cite{Stawarz_2019} the resolution factor is almost always below $0.3$. This resolution factor is thus comparable with the one we obtain for our simulations, in particular for SIM B-wf and SIM B-sf, for which it is $0.22$, in all directions (instead, we obtain $0.9$ for simulation run SIM A, corresponding to the $z$-direction which is the less resolved, while for the $x$ and $y$-direction we obtain $0.5$). Given these values, we have decided to accept all the previously selected data into a MCA procedure which is to be compared to that performed over HVM simulation results. 

By the whole procedure just detailed, a dataset containing about $200$ thousand points has been selected. Given these data, it is possible to obtain a valid statistic only for what we called ``generic'' points, since any further selection aiming to retain only samples retrieved inside current structures would leave us with no more than few thousand points, which are not sufficient for statistics in our case. 

\bibliographystyle{apalike}
\bibliography{main.bib}

\begin{thebibliography}{}

\bibitem[{Agudelo Rueda} et~al., 2021]{AgudeloRuedaJPP2021}
{Agudelo Rueda}, J.~A., {Verscharen}, D., {Wicks}, R.~T., {Owen}, C.~J.,
  {Nicolaou}, G., {Walsh}, A.~P., {Zouganelis}, I., {Germaschewski}, K., and
  {Vargas Dom{\'\i}nguez}, S. (2021).
\newblock {Three-dimensional magnetic reconnection in particle-in-cell
  simulations of anisotropic plasma turbulence}.
\newblock {\em Journal of Plasma Physics}, 87(3):905870228.

\bibitem[{Arzamasskiy} et~al., 2019]{ArzamasskiyAPJ2019}
{Arzamasskiy}, L., {Kunz}, M.~W., {Chandran}, B. D.~G., and {Quataert}, E.
  (2019).
\newblock {Hybrid-kinetic Simulations of Ion Heating in Alfv{\'e}nic
  Turbulence}.
\newblock {\em \apj}, 879(1):53.

\bibitem[Boldyrev and Perez, 2012]{boldyrev_spectrum_2012}
Boldyrev, S. and Perez, J.~C. (2012).
\newblock Spectrum of {Kinetic} {Alfven} {Turbulence}.
\newblock {\em The Astrophysical Journal}, 758(2):L44.
\newblock arXiv: 1204.5809.

\bibitem[Bruno et~al., 2001]{BRUNO20011201}
Bruno, R., Carbone, V., Veltri, P., Pietropaolo, E., and Bavassano, B. (2001).
\newblock Identifying intermittency events in the solar wind.
\newblock {\em Planetary and Space Science}, 49(12):1201--1210.
\newblock Nonlinear Dynamics and Fraactals in Space.

\bibitem[Burch et~al., 2016]{burch_magnetospheric_2016}
Burch, J.~L., Moore, T.~E., Torbert, R.~B., and Giles, B.~L. (2016).
\newblock Magnetospheric {Multiscale} {Overview} and {Science} {Objectives}.
\newblock {\em Space Science Reviews}, 199(1-4):5--21.

\bibitem[Califano et~al., 2020]{Califano_2020}
Califano, F., Cerri, S.~S., Faganello, M., Laveder, D., Sisti, M., and Kunz,
  M.~W. (2020).
\newblock Electron-only reconnection in plasma turbulence.
\newblock {\em Frontiers in Physics}, 8:317.

\bibitem[{Camporeale} et~al., 2018]{CamporealePRL2018}
{Camporeale}, E., {Sorriso-Valvo}, L., {Califano}, F., and {Retin{\`o}}, A.
  (2018).
\newblock {Coherent Structures and Spectral Energy Transfer in Turbulent
  Plasma: A Space-Filter Approach}.
\newblock {\em \prl}, 120(12):125101.

\bibitem[Carbone et~al., 1990]{carbone_coherent_1990}
Carbone, V., Veltri, P., and Mangeney, A. (1990).
\newblock Coherent structure formation and magnetic field line reconnection in
  magnetohydrodynamic turbulence.
\newblock {\em Physics of Fluids A: Fluid Dynamics}, 2(8):1487--1496.

\bibitem[Cerri and Califano, 2017]{cerri_reconnection_2017}
Cerri, S.~S. and Califano, F. (2017).
\newblock Reconnection and small-scale fields in {2D}-{3V} hybrid-kinetic
  driven turbulence simulations.
\newblock {\em New Journal of Physics}, 19(2):025007.

\bibitem[Cerri et~al., 2017a]{cerri_plasma_2017}
Cerri, S.~S., Franci, L., Califano, F., Landi, S., and Hellinger, P. (2017a).
\newblock Plasma turbulence at ion scales: a comparison between particle in
  cell and {Eulerian} hybrid-kinetic approaches.
\newblock {\em Journal of Plasma Physics}, 83(2):705830202.

\bibitem[{Cerri} et~al., 2019]{CerriFSPAS2019}
{Cerri}, S.~S., {Gro\v{s}elj}, D., and {Franci}, L. (2019).
\newblock {Kinetic plasma turbulence: recent insights and open questions from
  3D3V simulations}.
\newblock {\em Frontiers in Astronomy and Space Sciences}, 6:64.

\bibitem[Cerri et~al., 2018]{cerri_dual_2018}
Cerri, S.~S., Kunz, M.~W., and Califano, F. (2018).
\newblock Dual phase-space cascades in {3D} hybrid-{Vlasov}-{Maxwell}
  turbulence.
\newblock {\em The Astrophysical Journal}, 856(1):L13.
\newblock arXiv: 1802.06133.

\bibitem[Cerri et~al., 2017b]{cerri_kinetic_2017}
Cerri, S.~S., Servidio, S., and Califano, F. (2017b).
\newblock Kinetic {Cascade} in {Solar}-wind {Turbulence}: {3D3V}
  {Hybrid}-kinetic {Simulations} with {Electron} {Inertia}.
\newblock {\em The Astrophysical Journal}, 846(2):L18.

\bibitem[{Chen}, 2016]{ChenJPP2016}
{Chen}, C.~H.~K. (2016).
\newblock {Recent progress in astrophysical plasma turbulence from solar wind
  observations}.
\newblock {\em Journal of Plasma Physics}, 82(6):535820602.

\bibitem[{Comisso} and {Sironi}, 2019]{ComissoSironiAPJ2019}
{Comisso}, L. and {Sironi}, L. (2019).
\newblock {The Interplay of Magnetically Dominated Turbulence and Magnetic
  Reconnection in Producing Nonthermal Particles}.
\newblock {\em \apj}, 886(2):122.

\bibitem[De~Giorgio et~al., 2017]{degiorgio-2017-beltramization}
De~Giorgio, E., Servidio, S., and P., V. (2017).
\newblock Coherent structure formation through nonlinear interactions in 2d
  magnetohydrodynamic turbulence.
\newblock {\em Scientific Reports}, 7.

\bibitem[{Dong} et~al., 2018]{DongPRL2018}
{Dong}, C., {Wang}, L., {Huang}, Y.-M., {Comisso}, L., and {Bhattacharjee}, A.
  (2018).
\newblock {Role of the Plasmoid Instability in Magnetohydrodynamic Turbulence}.
\newblock {\em \prl}, 121(16):165101.

\bibitem[Dunlop et~al., 1988]{dunlop_analysis_1988}
Dunlop, M., Southwood, D., Glassmeier, K.-H., and Neubauer, F. (1988).
\newblock Analysis of multipoint magnetometer data.
\newblock {\em Advances in Space Research}, 8(9-10):273--277.

\bibitem[Fadanelli et~al., 2019]{fadanelli_fourspacecraft_2019}
Fadanelli, S., Lavraud, B., Califano, F., Jacquey, C., Vernisse, Y., Kacem, I.,
  Penou, E., Gershman, D.~J., Dorelli, J., Pollock, C., Giles, B.~L., Avanov,
  L.~A., Burch, J., Chandler, M.~O., Coffey, V.~N., Eastwood, J.~P., Ergun, R.,
  Farrugia, C.~J., Fuselier, S.~A., Genot, V.~N., Grigorenko, E., Hasegawa, H.,
  Khotyaintsev, Y., Le~Contel, O., Marchaudon, A., Moore, T.~E., Nakamura, R.,
  Paterson, W.~R., Phan, T., Rager, A.~C., Russell, C.~T., Saito, Y., Sauvaud,
  J., Schiff, C., Smith, S.~E., Toledo~Redondo, S., Torbert, R.~B., Wang, S.,
  and Yokota, S. (2019).
\newblock Four‐{Spacecraft} {Measurements} of the {Shape} and
  {Dimensionality} of {Magnetic} {Structures} in the {Near}‐{Earth} {Plasma}
  {Environment}.
\newblock {\em Journal of Geophysical Research: Space Physics},
  124(8):6850--6868.

\bibitem[Franci et~al., 2017]{franci_magnetic_2017}
Franci, L., Cerri, S.~S., Califano, F., Landi, S., Papini, E., Verdini, A.,
  Matteini, L., Jenko, F., and Hellinger, P. (2017).
\newblock Magnetic {Reconnection} as a {Driver} for a {Sub}-ion-scale {Cascade}
  in {Plasma} {Turbulence}.
\newblock {\em The Astrophysical Journal}, 850(1):L16.

\bibitem[{Gingell} et~al., 2020]{GingellJGRA2020}
{Gingell}, I., {Schwartz}, S.~J., {Eastwood}, J.~P., {Stawarz}, J.~E., {Burch},
  J.~L., {Ergun}, R.~E., {Fuselier}, S.~A., {Gershman}, D.~J., {Giles}, B.~L.,
  {Khotyaintsev}, Y.~V., {Lavraud}, B., {Lindqvist}, P.~A., {Paterson}, W.~R.,
  {Phan}, T.~D., {Russell}, C.~T., {Strangeway}, R.~J., {Torbert}, R.~B., and
  {Wilder}, F. (2020).
\newblock {Statistics of Reconnecting Current Sheets in the Transition Region
  of Earth's Bow Shock}.
\newblock {\em Journal of Geophysical Research (Space Physics)}, 125(1):e27119.

\bibitem[{Gosling} and {Phan}, 2013]{GoslingPhanAPJ2013}
{Gosling}, J.~T. and {Phan}, T.~D. (2013).
\newblock {Magnetic Reconnection in the Solar Wind at Current Sheets Associated
  with Extremely Small Field Shear Angles}.
\newblock {\em \apjl}, 763(2):L39.

\bibitem[{Greco} et~al., 2018]{GrecoSSRv2018}
{Greco}, A., {Matthaeus}, W.~H., {Perri}, S., {Osman}, K.~T., {Servidio}, S.,
  {Wan}, M., and {Dmitruk}, P. (2018).
\newblock {Partial Variance of Increments Method in Solar Wind Observations and
  Plasma Simulations}.
\newblock {\em \ssr}, 214(1):1.

\bibitem[Greco et~al., 2009]{Greco-2009-PhysRevE.80.046401}
Greco, A., Matthaeus, W.~H., Servidio, S., and Dmitruk, P. (2009).
\newblock Waiting-time distributions of magnetic discontinuities: Clustering or
  poisson process?
\newblock {\em Phys. Rev. E}, 80:046401.

\bibitem[{Gro{\v{s}}elj} et~al., 2017]{GroseljAPJ2017}
{Gro{\v{s}}elj}, D., {Cerri}, S.~S., {Ba{\~n}{\'o}n Navarro}, A., {Willmott},
  C., {Told}, D., {Loureiro}, N.~F., {Califano}, F., and {Jenko}, F. (2017).
\newblock {Fully Kinetic versus Reduced-kinetic Modeling of Collisionless
  Plasma Turbulence}.
\newblock {\em \apj}, 847(1):28.

\bibitem[{Khabarova} et~al., 2021]{KhabarovaSSRv2021}
{Khabarova}, O., {Malandraki}, O., {Malova}, H., {Kislov}, R., {Greco}, A.,
  {Bruno}, R., {Pezzi}, O., {Servidio}, S., {Li}, G., {Matthaeus}, W., {Le
  Roux}, J., {Engelbrecht}, N.~E., {Pecora}, F., {Zelenyi}, L., {Obridko}, V.,
  and {Kuznetsov}, V. (2021).
\newblock {Current Sheets, Plasmoids and Flux Ropes in the Heliosphere. Part I.
  2-D or not 2-D? General and Observational Aspects}.
\newblock {\em \ssr}, 217(3):38.

\bibitem[Loureiro and Boldyrev, 2017]{loureiro_collisionless_2017}
Loureiro, N.~F. and Boldyrev, S. (2017).
\newblock Collisionless {Reconnection} in {Magnetohydrodynamic} and {Kinetic}
  {Turbulence}.
\newblock {\em The Astrophysical Journal}, 850(2):182.

\bibitem[{Mallet} et~al., 2017]{MalletJPP2017}
{Mallet}, A., {Schekochihin}, A.~A., and {Chandran}, B.~D.~G. (2017).
\newblock Disruption of alfvénic turbulence by magnetic reconnection in a
  collisionless plasma.
\newblock {\em Journal of Plasma Physics}, 83(6).

\bibitem[{Matthaeus} et~al., 2020]{MatthaeusAPJ2020}
{Matthaeus}, W.~H., {Yang}, Y., {Wan}, M., {Parashar}, T.~N., {Bandyopadhyay},
  R., {Chasapis}, A., {Pezzi}, O., and {Valentini}, F. (2020).
\newblock {Pathways to Dissipation in Weakly Collisional Plasmas}.
\newblock {\em \apj}, 891(1):101.

\bibitem[Meyrand and Galtier, 2013]{meyrand_2013}
Meyrand, R. and Galtier, S. (2013).
\newblock Anomalous ${k}_{\ensuremath{\perp}}^{\ensuremath{-}8/3}$ spectrum in
  electron magnetohydrodynamic turbulence.
\newblock {\em Physical review letters}, 111:264501.

\bibitem[{Navarro} et~al., 2016]{BanonNavarroPRL2016}
{Navarro}, A.~B., {Teaca}, B., {Told}, D., {Groselj}, D., {Crandall}, P., and
  {Jenko}, F. (2016).
\newblock {Structure of Plasma Heating in Gyrokinetic Alfv{\'e}nic Turbulence}.
\newblock {\em \prl}, 117(24):245101.

\bibitem[{Osman} et~al., 2014]{OsmanPRL2014}
{Osman}, K.~T., {Matthaeus}, W.~H., {Gosling}, J.~T., {Greco}, A., {Servidio},
  S., {Hnat}, B., {Chapman}, S.~C., and {Phan}, T.~D. (2014).
\newblock {Magnetic Reconnection and Intermittent Turbulence in the Solar
  Wind}.
\newblock {\em \prl}, 112(21):215002.

\bibitem[Osman et~al., 2012]{Osman-2012-PhysRevLett.108.261103}
Osman, K.~T., Matthaeus, W.~H., Hnat, B., and Chapman, S.~C. (2012).
\newblock Kinetic signatures and intermittent turbulence in the solar wind
  plasma.
\newblock {\em Phys. Rev. Lett.}, 108:261103.

\bibitem[Papini et~al., 2019]{papini_can_2019}
Papini, E., Franci, L., Landi, S., Verdini, A., Matteini, L., and Hellinger, P.
  (2019).
\newblock Can {Hall} {Magnetohydrodynamics} explain plasma turbulence at
  sub-ion scales?
\newblock {\em The Astrophysical Journal}, 870(1):52.
\newblock arXiv: 1810.02210.

\bibitem[{Passot} et~al., 2014]{PassotEPJD2014}
{Passot}, T., {Henri}, P., {Laveder}, D., and {Sulem}, P.-L. (2014).
\newblock {Fluid simulations of ion scale plasmas with weakly distorted
  magnetic fields. FLR-Landau fluid simulations}.
\newblock {\em European Physical Journal D}, 68(7):207.

\bibitem[{Pecora} et~al., 2019]{PecoraAPJ2019}
{Pecora}, F., {Greco}, A., {Hu}, Q., {Servidio}, S., {Chasapis}, A.~G., and
  {Matthaeus}, W.~H. (2019).
\newblock {Single-spacecraft Identification of Flux Tubes and Current Sheets in
  the Solar Wind}.
\newblock {\em \apjl}, 881(1):L11.

\bibitem[Phan et~al., 2018]{phan_electron_2018}
Phan, T.~D., Eastwood, J.~P., Shay, M.~A., Drake, J.~F., Sonnerup, B. U.~O.,
  Fujimoto, M., Cassak, P.~A., Øieroset, M., Burch, J.~L., Torbert, R.~B.,
  Rager, A.~C., Dorelli, J.~C., Gershman, D.~J., Pollock, C., Pyakurel, P.~S.,
  Haggerty, C.~C., Khotyaintsev, Y., Lavraud, B., Saito, Y., Oka, M., Ergun,
  R.~E., Retino, A., Le~Contel, O., Argall, M.~R., Giles, B.~L., Moore, T.~E.,
  Wilder, F.~D., Strangeway, R.~J., Russell, C.~T., Lindqvist, P.~A., and
  Magnes, W. (2018).
\newblock Electron magnetic reconnection without ion coupling in {Earth}’s
  turbulent magnetosheath.
\newblock {\em Nature}, 557(7704):202--206.

\bibitem[{Podesta}, 2017]{PodestaJGRA2017}
{Podesta}, J.~J. (2017).
\newblock {The most intense current sheets in the high-speed solar wind near 1
  AU}.
\newblock {\em Journal of Geophysical Research (Space Physics)},
  122(3):2795--2823.

\bibitem[Pollock et~al., 2016]{pollock_fast_2016}
Pollock, C., Moore, T., Jacques, A., Burch, J., Gliese, U., Saito, Y., Omoto,
  T., Avanov, L., Barrie, A., Coffey, V., Dorelli, J., Gershman, D., Giles, B.,
  Rosnack, T., Salo, C., Yokota, S., Adrian, M., Aoustin, C., Auletti, C.,
  Aung, S., Bigio, V., Cao, N., Chandler, M., Chornay, D., Christian, K.,
  Clark, G., Collinson, G., Corris, T., De~Los~Santos, A., Devlin, R., Diaz,
  T., Dickerson, T., Dickson, C., Diekmann, A., Diggs, F., Duncan, C.,
  Figueroa-Vinas, A., Firman, C., Freeman, M., Galassi, N., Garcia, K.,
  Goodhart, G., Guererro, D., Hageman, J., Hanley, J., Hemminger, E., Holland,
  M., Hutchins, M., James, T., Jones, W., Kreisler, S., Kujawski, J., Lavu, V.,
  Lobell, J., LeCompte, E., Lukemire, A., MacDonald, E., Mariano, A., Mukai,
  T., Narayanan, K., Nguyan, Q., Onizuka, M., Paterson, W., Persyn, S.,
  Piepgrass, B., Cheney, F., Rager, A., Raghuram, T., Ramil, A., Reichenthal,
  L., Rodriguez, H., Rouzaud, J., Rucker, A., Saito, Y., Samara, M., Sauvaud,
  J.-A., Schuster, D., Shappirio, M., Shelton, K., Sher, D., Smith, D., Smith,
  K., Smith, S., Steinfeld, D., Szymkiewicz, R., Tanimoto, K., Taylor, J.,
  Tucker, C., Tull, K., Uhl, A., Vloet, J., Walpole, P., Weidner, S., White,
  D., Winkert, G., Yeh, P.-S., and Zeuch, M. (2016).
\newblock Fast {Plasma} {Investigation} for {Magnetospheric} {Multiscale}.
\newblock {\em Space Science Reviews}, 199(1-4):331--406.

\bibitem[Russell et~al., 2015]{russell_magnetic_2015}
Russell, A. J.~B., Yeates, A.~R., and Eastwood, J.~P. (2015).
\newblock Magnetic reconnection now and in the future.
\newblock {\em Astronomy \& Geophysics}, 56(6):6.18--6.23.

\bibitem[{Sahraoui} et~al., 2020]{SahraouiRMPP2020}
{Sahraoui}, F., {Hadid}, L., and {Huang}, S. (2020).
\newblock {Magnetohydrodynamic and kinetic scale turbulence in the near-Earth
  space plasmas: a (short) biased review}.
\newblock {\em Review of Modern Plasma Physics}, 4(1):4.

\bibitem[{Schekochihin} et~al., 2009]{SchekochihinAPJS2009}
{Schekochihin}, A.~A., {Cowley}, S.~C., {Dorland}, W., {Hammett}, G.~W.,
  {Howes}, G.~G., {Quataert}, E., and {Tatsuno}, T. (2009).
\newblock {Astrophysical Gyrokinetics: Kinetic and Fluid Turbulent Cascades in
  Magnetized Weakly Collisional Plasmas}.
\newblock {\em The Astrophysical Journal Supplement Series}, 182:310--377.

\bibitem[Servidio et~al., 2009]{servidio_magnetic_2009}
Servidio, S., Matthaeus, W.~H., Shay, M.~A., Cassak, P.~A., and Dmitruk, P.
  (2009).
\newblock Magnetic {Reconnection} in {Two}-{Dimensional} {Magnetohydrodynamic}
  {Turbulence}.
\newblock {\em Physical Review Letters}, 102(11):115003.

\bibitem[Shen et~al., 2007]{shen_new_2007}
Shen, C., Dunlop, M., Li, X., Liu, Z.~X., Balogh, A., Zhang, T.~L., Carr,
  C.~M., Shi, Q.~Q., and Chen, Z.~Q. (2007).
\newblock New approach for determining the normal of the bow shock based on
  {Cluster} four-point magnetic field measurements: {NORMAL} {OF} {EARTH}'{S}
  {BOW} {SHOCK}.
\newblock {\em Journal of Geophysical Research: Space Physics},
  112(A3):n/a--n/a.

\bibitem[Shi et~al., 2005]{shi_dimensional_2005}
Shi, Q.~Q., Shen, C., Pu, Z.~Y., Dunlop, M.~W., Zong, Q.-G., Zhang, H., Xiao,
  C.~J., Liu, Z.~X., and Balogh, A. (2005).
\newblock Dimensional analysis of observed structures using multipoint magnetic
  field measurements: {Application} to {Cluster}: {STRUCTURE} {DIMENSIONALITY}
  {DETERMINATION}.
\newblock {\em Geophysical Research Letters}, 32(12):n/a--n/a.

\bibitem[Shi et~al., 2019]{shi_dimensionality_2019}
Shi, Q.~Q., Tian, A.~M., Bai, S.~C., Hasegawa, H., Degeling, A.~W., Pu, Z.~Y.,
  Dunlop, M., Guo, R.~L., Yao, S.~T., Zong, Q.-G., Wei, Y., Zhou, X.-Z., Fu,
  S.~Y., and Liu, Z.~Q. (2019).
\newblock Dimensionality, {Coordinate} {System} and {Reference} {Frame} for
  {Analysis} of {In}-{Situ} {Space} {Plasma} and {Field} {Data}.
\newblock {\em Space Science Reviews}, 215(4):35.

\bibitem[Sorriso-Valvo et~al., 2018]{sorriso-2018}
Sorriso-Valvo, L., Carbone, F., Perri, S., Greco, A., Marino, R., and Bruno, R.
  (2018).
\newblock On the statistical properties of turbulent energy transfer rate in
  the inner heliosphere.
\newblock {\em Solar Physics}, 293.

\bibitem[Stawarz et~al., 2019]{Stawarz_2019}
Stawarz, J.~E., Eastwood, J.~P., Phan, T.~D., Gingell, I.~L., Shay, M.~A.,
  Burch, J.~L., Ergun, R.~E., Giles, B.~L., Gershman, D.~J., Contel, O.~L.,
  Lindqvist, P.-A., Russell, C.~T., Strangeway, R.~J., Torbert, R.~B., Argall,
  M.~R., Fischer, D., Magnes, W., and Franci, L. (2019).
\newblock Properties of the turbulence associated with electron-only magnetic
  reconnection in earth's magnetosheath.
\newblock {\em The Astrophysical Journal}, 877(2):L37.

\bibitem[{Sulem} et~al., 2016]{SulemAPJ2016}
{Sulem}, P.~L., {Passot}, T., {Laveder}, D., and {Borgogno}, D. (2016).
\newblock {Influence of the Nonlinearity Parameter on the Solar Wind Sub-ion
  Magnetic Energy Spectrum: FLR-Landau Fluid Simulations}.
\newblock {\em \apj}, 818:66.

\bibitem[{TenBarge} and {Howes}, 2013]{TenBargeHowesAPJ2013}
{TenBarge}, J.~M. and {Howes}, G.~G. (2013).
\newblock {Current Sheets and Collisionless Damping in Kinetic Plasma
  Turbulence}.
\newblock {\em \apjl}, 771(2):L27.

\bibitem[{Told} et~al., 2015]{ToldPRL2015}
{Told}, D., {Jenko}, F., {TenBarge}, J.~M., {Howes}, G.~G., and {Hammett},
  G.~W. (2015).
\newblock {Multiscale Nature of the Dissipation Range in Gyrokinetic
  Simulations of Alfv{\'e}nic Turbulence}.
\newblock {\em PhRvL}, 115(2):025003.

\bibitem[Uritsky et~al., 2010]{uritsky_2010}
Uritsky, V., Pouquet, A., Rosenberg, D., Mininni, P., and Donovan, E. (2010).
\newblock Structures in magnetohydrodynamic turbulence: Detection and scaling.
\newblock {\em Physical review. E, Statistical, nonlinear, and soft matter
  physics}, 82:056326.

\bibitem[Valentini et~al., 2007]{valentini_hybrid-vlasov_2007}
Valentini, F., Trávníček, P., Califano, F., Hellinger, P., and Mangeney, A.
  (2007).
\newblock A hybrid-{Vlasov} model based on the current advance method for the
  simulation of collisionless magnetized plasma.
\newblock {\em Journal of Computational Physics}, 225(1):753--770.

\bibitem[{Vech} et~al., 2018]{VechAPJL2018}
{Vech}, D., {Mallet}, A., {Klein}, K.~G., and {Kasper}, J.~C. (2018).
\newblock {Magnetic Reconnection May Control the Ion-scale Spectral Break of
  Solar Wind Turbulence}.
\newblock {\em \apjl}, 855(2):L27.

\bibitem[{Zhdankin} et~al., 2014]{ZhdankinAPJ2014}
{Zhdankin}, V., {Boldyrev}, S., {Perez}, J.~C., and {Tobias}, S.~M. (2014).
\newblock {Energy Dissipation in Magnetohydrodynamic Turbulence: Coherent
  Structures or ``Nanoflares''?}
\newblock {\em \apj}, 795(2):127.

\bibitem[Zhdankin et~al., 2013]{zhdankin_statistical_2013}
Zhdankin, V., Uzdensky, D.~A., Perez, J.~C., and Boldyrev, S. (2013).
\newblock Statistical {Analysis} of {Current} {Sheets} in {Three}-{Dimensional}
  {Magnetohydrodynamic} {Turbulence}.
\newblock {\em The Astrophysical Journal}, 771(2):124.
\newblock arXiv: 1302.1460.

\bibitem[{Zhdankin} et~al., 2017]{ZhdankinPRL2017}
{Zhdankin}, V., {Werner}, G.~R., {Uzdensky}, D.~A., and {Begelman}, M.~C.
  (2017).
\newblock {Kinetic Turbulence in Relativistic Plasma: From Thermal Bath to
  Nonthermal Continuum}.
\newblock {\em \prl}, 118(5):055103.

\end{thebibliography}

\end{document}